\documentclass{aa}  

\usepackage{booktabs}
\usepackage{siunitx}
\usepackage{makecell}
\usepackage{graphicx}
\usepackage[colorlinks=true,allcolors=blue]{hyperref}
\bibpunct{(}{)}{;}{a}{}{,} 

\usepackage{txfonts}
\newcommand{\CIV}{$\mathrm{C}\textsc{ IV}$\xspace}
\newcommand{\CII}{$[\mathrm{C}\textsc{ ii}]$\xspace}

\newcommand{\HeIIline}{$\mathrm{He}\textsc{ ii}\,\lambda 1640$\xspace}
\newcommand{\Ha}{$\mathrm{H}\alpha$\xspace}
\newcommand{\Hb}{$\mathrm{H}\beta$\xspace}
\newcommand{\Hg}{$\mathrm{H}\gamma$\xspace}

\newcommand{\OII}{$[\mathrm{O}\textsc{ ii}]$\xspace}
\newcommand{\OIIlines}{$[\mathrm{O}\textsc{ ii}]\,\lambda\lambda 3726,29$\xspace}

\newcommand{\NeIIIline}{$[\mathrm{Ne}\textsc{ iii}]\,\lambda 3869$\xspace}
\newcommand{\OIII}{$[\mathrm{O}\textsc{ iii}]$\xspace}

\newcommand{\OIIIdm}{$[\mathrm{O}\textsc{ iii}]\,\lambda 4363$\xspace}
\newcommand{\OIIIb}{$[\mathrm{O}\textsc{ iii}]\,\lambda 5007$\xspace}

\newcommand{\NII}{$[\mathrm{N}\textsc{ ii}]$\xspace}

\newcommand{\NIIb}{$[\mathrm{N}\textsc{ ii}]\,\lambda 6584$\xspace}
\newcommand{\SII}{$[\mathrm{S}\textsc{ ii}]$\xspace}
\newcommand{\SIIlines}{$[\mathrm{S}\textsc{ ii}]\,\lambda\lambda 6716,31$\xspace}

\newcommand{\SIIIline}{$[\mathrm{S}\textsc{ iii}]\,\lambda 9532$\xspace}

\begin{document} 

   \title{GA-NIFS: High prevalence of dusty and metal-enriched outflows in massive and luminous star-forming galaxies at $z\sim3-9$ }

 \titlerunning{GA-NIFS: Spatially resolved outflows at $3<z<9$}
   \author{B. Rodr\'iguez Del Pino
          \inst{\ref{iCAB}}\thanks{e-mail: brodriguez@cab.inta-csic.es}
          \and
          S. Arribas\inst{\ref{iCAB}}
          \and
          M. Perna\inst{\ref{iCAB}}
          \and    
          I. Lamperti\inst{\ref{iUNIFI}, \ref{iOAA}}
          \and
          A. Bunker\inst{\ref{iOxf}}
          \and
          S. Carniani\inst{\ref{iNorm}}
          \and 
          S. Charlot\inst{\ref{iSor}}
          \and
          F. D'Eugenio\inst{\ref{iKav},\ref{iCav}}  
          \and
          R. Maiolino\inst{\ref{iKav},\ref{iCav}, \ref{iUCL}}            \and
          H. \"{U}bler\inst{\ref{iMPE}} 
          \and 
          E. Bertola\inst{\ref{iOAA}}
          \and 
          T. B{\"o}ker\inst{\ref{iESOba}}
          \and 
          G. Cresci\inst{\ref{iOAA}}        
          \and
          G. C. Jones\inst{\ref{iKav},\ref{iCav}}         
          \and          
          C. Marconcini\inst{\ref{iUNIFI}, \ref{iOAA}}
          \and
          E. Parlanti\inst{\ref{iNorm}}
          \and 
          J. Scholtz\inst{\ref{iKav},\ref{iCav}}
          \and 
          G. Venturi\inst{\ref{iNorm}}
          \and  
          S. Zamora \inst{\ref{iNorm}}
          }

   \institute{
            Centro de Astrobiolog\'ia (CAB), CSIC--INTA, Cra. de Ajalvir Km.~4, 28850 -- Torrej\'on de Ardoz, Madrid, Spain\label{iCAB}
    \and
            Università di Firenze, Dipartimento di Fisica e Astronomia, via G. Sansone 1, 50019 Sesto F.no, Firenze, Italy\label{iUNIFI}
    \and
            INAF - Osservatorio Astrofisico di Arcetri, Largo E. Fermi 5, I-50125 Firenze, Italy\label{iOAA}
    \and
            Department of Physics, University of Oxford, Denys Wilkinson Building, Keble Road, Oxford OX1 3RH, UK\label{iOxf}
    \and        
            European Space Agency, c/o STScI, 3700 San Martin Drive, Baltimore, MD 21218, USA\label{iESOba}
    \and
            Scuola Normale Superiore, Piazza dei Cavalieri 7, I-56126 Pisa, Italy\label{iNorm}   
    \and 
            Sorbonne Universit\'e, CNRS, UMR 7095, Institut d’Astrophysique de Paris, 98 bis bd Arago, 75014 Paris, France\label{iSor} 
    \and
            Max-Planck-Institut f\"ur extraterrestrische Physik (MPE), Gie{\ss}enbachstra{\ss}e 1, 85748 Garching, Germany\label{iMPE}
    \and
            Kavli Institute for Cosmology, University of Cambridge, Madingley Road, Cambridge, CB3 0HA, UK\label{iKav}
    \and
            Cavendish Laboratory - Astrophysics Group, University of Cambridge, 19 JJ Thomson Avenue, Cambridge, CB3 0HE, UK\label{iCav}
    \and
            Department of Physics and Astronomy, University College London, Gower Street, London WC1E 6BT, UK\label{iUCL}            
   }

   \date{Received ... ; accepted ...}

  \abstract 
{
  We present the search and characterisation of ionised outflows in a sample of 15 star-forming systems at $z\sim3-9$ with no evidence of active galactic nuclei (AGNs) observed with the JWST/NIRSpec instrument in Integral Field Spectroscopic mode (IFS) as part of the `Galaxy Assembly with NIRSpec IFS' (GA-NIFS) program. Some of the target systems are comprised of various individual galaxies, adding up to a total of $40$ individual objects. Our sample encompasses the high mass end of the galaxy populations, with most of the sample having stellar masses within log$_{10}$~(M$_\star$/M$_\odot$)~=~$9.5-11$, while previous studies on high-z star formation driven outflows generally contain galaxies of log$_{10}$~(M$_\star$/M$_\odot$)$<9.5$. Leveraging the spatially resolved information and rest-frame optical coverage provided by NIRSpec IFS data, we analysed the \OIIIb and \Ha emission lines to search for broad kinematic components associated with galactic outflows. Crucially, the IFS observations allowed us to directly isolate the regions hosting the outflows, rather than relying on integrated spectra. We identified signatures of ionised outflows in 16 individual galaxies/regions (in 13 out of 15 systems), although we consider two of them only as `candidates', as they could be related to mergers or tidal interactions. After constraining their spatial location and extent, we characterised the integrated properties of the outflowing gas and evaluated the impact on the host galaxies. We find that the outflowing gas is more dust attenuated (by $A_{\rm V}\sim0.59$~mag on average) and metal enriched ($\sim0.13$~dex) than the interstellar medium of the host galaxies, suggesting that outflows contribute to distributing dust and metals around them. The outflows identified in this study present velocity dispersions within $\sigma_{\rm out}\sim130-340$~km~s$^{-1}$ and outflow velocities V$_{\rm out}\sim170-600$~km~s$^{-1}$. Considering also less luminous and less massive star-forming galaxies from previous works, there is a statistically significant correlation between outflow velocity ($v_{\rm out}$) and star formation rate (SFR). The generally low mass-loading factors ($\eta=\dot{\rm M}_{\rm out}/SFR$~$<1$, in nine out of 14 outflows) obtained for the ionised outflows suggest that they do not suppress star formation in the host galaxies. Moreover, their velocities are not high enough to escape their hosts and reach the circumgalactic medium. Our results indicate that ejective feedback through ionised outflows is inefficient in high-mass and luminous star-forming galaxies within the first 2 Gyr of cosmic time.
}


   \maketitle
%

\section{Introduction}

Galactic outflows triggered by intense episodes of star formation or by the accretion of material onto active galactic nuclei (AGNs) are widely regarded as fundamental drivers of galaxy evolution across cosmic time. They are believed to regulate and suppress both star formation and black hole growth and are considered the main mechanism through which dust and metals are redistributed on galactic scales or expelled into the intergalactic medium. Due to their importance, the identification and characterisation of galactic outflows are among the most extensively studied topics in galaxy evolution, both in the local Universe and at high redshift \citep[see reviews by][]{rupkeReviewRecentObservations2018, veilleuxCoolOutflowsGalaxies2020}.

Outflows are observed across multiple gas phases (ionised, neutral, and molecular), all of which should be taken into account to fully characterise their global properties and impact on host galaxies. Nevertheless, the ionised phase is the most frequently studied, as it can be traced through strong rest-frame optical emission lines such as \Ha and \OIIIb. Since these lines can be accessed by ground-based observatories with spectroscopic capabilities out to $z\leq2.5$ (where they are shifted into the near-infrared regime), ionised outflows have been extensively studied in large samples of star-forming galaxies and AGNs in the local Universe \citep[e.g.][]{ciconeOutflowsComplexStellar2016, pernaXraySDSSSample2017} and across the past $\sim11$~Gyrs of cosmic time \citep[e.g.][]{newmanSINSZCSINFSURVEY2012, forsterschreiberKMOS3DSurveyDemographics2019b, swinbankEnergeticsStarburstdrivenOutflows2019a, concasBeingKLEVERCosmic2022}. Overall, ionised galactic outflows are frequently ($70-100\%$) identified in AGN hosts both at low \citep[][]{harrisonKiloparsecscaleOutflowsAre2014, villarmartinTriggeringMechanismProperties2014} and high ($z\sim3$) redshifts \citep[][]{brusaXshooterRevealsPowerful2015, templeIIIEmissionLine2019, kakkadSUPERIISpatially2020, tozziSUPERVIIIFast2024}. In the local Universe, the incidence of outflows in main-sequence star-forming galaxies is very low \citep[e.g. $1\%$;][]{rodriguezdelpinoPropertiesionizedOutflows2019}, but since their incidence increases with star formation activity, they become almost ubiquitous in galaxies with the highest star formation rates (SFRs), such as ultra-luminous infrared galaxies \citep[ULIRGs; e.g.][]{arribasionizedGasOutflows2014}. At cosmic noon ($z \sim 2-3$), the detection rate in star-forming galaxies increases, probably because observations generally target brighter (and thus more active and massive) systems \citep{genzelSinsSurvey22011, newmanSINSZCSINFSURVEY2012, forsterschreiberKMOS3DSurveyDemographics2019b, freemanMOSDEFSurveyBroad2019}. In fact, their incidence seems to significantly decrease in low-mass systems \citep[log$_{10}$(M$_\star$/M$_\odot$)~$<$~9.6;][]{concasBeingKLEVERCosmic2022}. In star-forming galaxies (i.e. galaxies without evidence of AGN activity), different works have found correlations between outflow kinematics and host galaxy properties such as stellar mass (M$_\star$) and SFR surface density ($\Sigma_{\rm SFR}$) at both low \citep[e.g.][]{averyIncidenceScalingRelations2021a} and high redshift \citep[e.g.][]{daviesKiloparsecScaleProperties2019, guptaMOSELSurveyExtremely2023, llerenaionizedGasKinematics2023}. Although outflows in star-forming galaxies can reach velocities of several hundred kilometres per second and entrain substantial amounts of ionised gas, their impact on the star formation activity of the hosts appears to be limited across cosmic time. This result is reflected in the low ratio between the mass-loss rate and the SFR in the host, also known as the mass-loading factor, $\eta$. At least in local galaxies, outflows do seem to influence the distribution of metals and dust. They are often found to be more dust-obscured and metal-enriched than their hosts \citep{rodriguezdelpinoPropertiesionizedOutflows2019}, potentially contributing to the shaping of the mass–metallicity relation \citep{chisholmMetalenrichedGalacticOutflows2018a}. At higher redshifts, however, this is a result that remains to be explored. 

Despite the extensive studies conducted up to $z\sim3$, identifying and characterising ionised outflows at earlier epochs has remained out of reach until recently, largely because no facilities offered spectroscopic coverage in the near-infrared regime beyond 2.5~$\mu$m, where the \Ha line is redshifted at $z\geq3$. The commissioning and successful launch of the James Webb Space Telescope (JWST) have opened a new era in the exploration of distant galaxies. In fact, several works have already started to characterise ionised outflows in star-forming galaxies at $3<z<9$ \citep{carnianiJADESIncidenceRate2024, zhangStatisticsGalaxyOutflows2024, cooperHighvelocityOutflowsOIII2025, crespo-gomezRIOJADustyOutflows2025, marques-chavesExtremelyUVbrightStarbursts2025, renRIOJAYoungStarburst2025,  xuStellarAGNDrivenOutflows2025a, zhuSystematicSearchGalaxies2025}. These studies report an incidence of ionised outflows that varies between 20-40\% \citep{carnianiJADESIncidenceRate2024, xuStellarAGNDrivenOutflows2025a} and $3.4\%$ \citep{cooperHighvelocityOutflowsOIII2025}. 
Such discrepancies are probably due to the application of stricter outflow selection criteria and to the different spectral resolutions of the observations employed in each work, since a higher spectral resolution favours the identification of a broad component. The studies have also reported contrasting results on the impact that star formation--driven outflows have on the star formation activity in the host galaxy (as quantified by the mass-loading factor, $\eta$) and on whether the outflowing gas reaches velocities sufficient to escape the host systems. Within similar ranges of stellar masses (log$_{10}$(M$_\star$/M$_\odot$)~=~$[7.5,9.5]$) and star formation activity (log$_{10}$(SFR)~=~$[-0.2,2]$ M$_\odot$yr$^{-1}$), \citet{carnianiJADESIncidenceRate2024} estimates median values of $\eta$\,$\sim$\,2 and outflow velocities that are on average three times higher than the escape velocity, while \cite{cooperHighvelocityOutflowsOIII2025} and \cite{xuStellarAGNDrivenOutflows2025a} find much lower values of $\eta$ ($0.1-1$) and outflow velocities not large enough to escape from the gravitational potential of the hosts. 

Although these previous JWST studies of ionised outflows at $3<z<9$ provide relevant information about their incidence, properties, and impact on the host galaxies, they are all based on integrated information obtained from Multi Object Spectroscopy (MOS) or Wide Field Slitless Spectroscopy (WFSS). Moreover, they have targeted galaxies with log~(M$_\star$/M$_\odot$) $<9.5$, while the higher mass range of starbursts, a regime where outflows are theorised to be more common but are increasingly suppressed by the deepening gravitational potential, remains to be explored. In this context, NIRSpec Integral Field Spectroscopy (IFS) observations provide, for the first time, the possibility to spatially resolve ionised outflows at high redshift. In fact, some studies have already started to identify and characterise spatially resolved ionised outflows in star-forming galaxies at $z>3$ \citep[][]{lampertiGANIFSJWSTNIRSpec2024a, rodriguezdelpinoGANIFSCoevolutionHighly2024, iveyExploringSpatiallyResolved2026, parlantiGANIFSMultiphaseAnalysis2025, marconciniGANIFSDissectingMultiple2025, scholtzGANIFSISMProperties2025, zamoraGANIFSUnderstandingionization2025}. In this work, we extend previous studies by leveraging the unique spatially resolved power of NIRSpec IFS observations of a relatively large sample of massive and luminous star-forming galaxies at $3<z<9$, delivering a benchmark study of star formation--driven ionised outflows in the early Universe.

This paper is organised as follows. In Sect.~\ref{sec:data} we describe the NIRSpec observations, the reduction, and the analysis of the data. In Sect.~\ref{sec:analysis} we describe the data analysis, including the spectral modelling of emission lines using the R2700 data and the stellar mass estimation using the R100 data. Section~\ref{sec:general_properties} contains a description of the general properties of the sample, and in Sect.~\ref{sec:outflow_regions} we describe the spectral modelling and the criteria to identify outflows. In Sect.~\ref{sec:results} we present the main results obtained, while a discussion on their implications is presented in Sect.~\ref{sec:Discussion}. Finally, a summary of the main results and our concluding remarks are presented in Sect.~\ref{sec:conclusions}.

Throughout this paper, we adopt a \citet{chabrierGalacticStellarSubstellar2003} initial mass function (IMF; $0.1-100$ M$_{\odot}$) and a standard $\Lambda$\ Cold Dark Matter (CDM) cosmology with $H_{\rm 0}$~=~70~km~s$^{\rm -1}$~Mpc$^{\rm -1}$, $\Omega_{\rm \Lambda}$~=~0.7, and $\Omega_{\rm m}$~=~0.3. Emission lines are referred to using their rest-frame air wavelengths, although for the analysis we used their vacuum wavelengths.

\section{Galaxy sample, observations, and data reduction}
\label{sec:data}

\subsection{GA-NIFS sample of star-forming galaxies}
The data presented in this paper are part of the JWST Cycle 1 observations obtained within the NIRSpec IFS Guaranteed Time  Observations (GTO) program `Galaxy Assembly with NIRSpec IFS' (GA-NIFS\footnote{https://ga-nifs.github.io/}; PIs: S. Arribas, R. Maiolino). This program consists of more than $300$~hours of NIRSpec/IFS observations of a sample of $>50$ galaxy systems at $z\sim2-11$, including quiescent \citep{deugenioFastrotatorPoststarburstGalaxy2024a, perez-gonzalezAcceleratedQuenchingChemical2025}, star-forming \citep{arribasGANIFSCoreExtremely2024a,  jonesGANIFSInterstellarMedium2026, jonesGANIFSJWSTNIRSpec2024,  lampertiGANIFSJWSTNIRSpec2024a, rodriguezdelpinoGANIFSCoevolutionHighly2024, jonesGANIFSWitnessingComplex2025, parlantiGANIFSMultiphaseAnalysis2025, marconciniGANIFSDissectingMultiple2025, scholtzGANIFSISMProperties2025} and AGN galaxies \citep{marshallGANIFSBlackHole2023, pernaGANIFSUltradenseInteracting2023, ublerGANIFSMassiveBlack2023c,parlantiGANIFSEarlystageFeedback2024b,pernaGANIFSGalaxywideOutflow2025, ublerGANIFSNIRSpecReveals2024a,ublerGANIFSJWSTDiscovers2024,bertolaGANIFSMapping352025, marshallGANIFSEIGERMerging2025, pernaGANIFSHighNumber2025,trefoloniGANIFSExtendedOIII2025,zamoraGANIFSHighlyOverdense2025}. The GA-NIFS sub-sample of star-forming galaxies was selected with the goal of exploring the high-mass end of the star formation main sequence as well as bright starbursts at early epochs, favouring the selection of extended objects and/or the presence of nearby companions, and therefore the sub-sample is appropriate for IFS studies and complementary to MOS survey samples \citep[e.g.][]{bunkerJADESNIRSpecInitial2024, finkelsteinCosmicEvolutionEarly2025, masedaNIRSpecWideGTO2024}. The sample of star-forming galaxies was also chosen based on the absence of clear X-ray emission and very broad emission lines that could be associated with the presence of an active supermassive black hole, based on pre-JWST observations. However, the classification of some of the formerly considered star-forming galaxies has been recently updated through the analysis of JWST NIRSpec/IFS data targeting the optical rest-frame emission that have revealed the presence of AGN activity (\citealt{ublerGANIFSMassiveBlack2023c, ublerGANIFSNIRSpecReveals2024a, pernaGANIFSHighNumber2025, venturiGANIFSPowerfulFrequent2025}). 

After removing the previously identified AGNs, we end up with a sample of 15 star-forming systems distributed in the redshift range $\sim3-9$. The individual properties of several of these systems have been already explored as part of the GA-NIFS survey, finding clear signatures of ionised outflows in some of them \citep[e.g.][]{lampertiGANIFSJWSTNIRSpec2024a, rodriguezdelpinoGANIFSCoevolutionHighly2024, parlantiGANIFSMultiphaseAnalysis2025, zamoraGANIFSUnderstandingionization2025} and tentative evidence in others \citep[e.g.][]{jonesGANIFSInterstellarMedium2026, jonesGANIFSJWSTNIRSpec2024}. 
In this work we provide a systematic and homogeneous identification and characterisation of ionised outflows in these previously studied star-forming galaxies and present, for the first time, an analysis of ionised outflows in three additional galaxies from GA-NIFS (GS\_23170, GS\_12306 and MACSJ0416-Y1).

\subsection{JWST/NIRSpec IFS observations}
The targets in our sample were observed with NIRSpec/IFS at high (R2700) and low spectral resolution (R100/PRISM) configurations, with the exception of COS\_B14$-$65666 (hereafter B14) that was only observed at medium-resolution (R1000). Observations were included in proposals  \#1208 (PI: C. Willott), \#1216, \#1217, \#1262 (PI: N. Luetzgendorf), and \#1264 (PI: L. Colina). The associated proposal IDs as well as the total exposure times for each galaxy are listed in Table~\ref{table:summary_observations}. Observations were executed applying a medium (0.5\arcsec) cycling pattern of eight dithers, covering a contiguous area of 3.1\arcsec~$\times~$3.2\arcsec{} with a native spaxel size of 0.1\arcsec{} \citep{bokerNearInfraredSpectrographNIRSpec2022, rigbySciencePerformanceJWST2023}. R100 observations provide a spectral resolution $R\sim30-330$ within $0.6\mu$m~$-$~$5.3\mu$m, R1000 G395M/F290LP observations provide $R\sim700-1200$ within $2.88\mu$m~$-$~$5.20\mu$m, and R2700 observations provide $R\sim1900-3500$ within $1.7\mu$m~$-$~$3.15\mu$m and $2.88\mu$m~$-$~$5.20\mu$m for G235H/F170LP and G395H/290LP, respectively. 

\begin{table*}
\caption{GA-NIFS targets included in this study.}
\centering
\small 
\begin{tabular}{llccccccc}
\toprule
\makecell{Target} & 
\makecell{Galaxies /\\ regions} & 
\makecell{Pointing \\ R.A. \\ (deg)} & 
\makecell{Pointing \\ Dec. \\ (deg)} & 
\makecell{Program} & 
\makecell{Redshift} & 
\makecell{T$_{\rm exp}$ (s) \\ (R2700 | R100)} &
\makecell{log$_{10}$M$_{\star}$ \\ (M$_\odot$)} & 
\makecell{log$_{10}SFR$ \\ (M$_\odot$yr$^{-1}$)} \\ 
\midrule
GS\_23170        & N, S           & 53.15746  & $-27.70911$ & 1216 & 2.979 & 14705.6 \,|\, 3968.2       & $9.69$   & $1.70\pm0.01$ \\
GS\_5001         & C              & 53.09725  & $-27.86588$ & 1216 & 3.471 & 14705.6 \,|\, 3968.2       & $10.89$  & $2.02\pm0.01$ \\
                 & N              &           &             &      & 3.474 &                            & $9.86$   & $1.77\pm0.02$ \\
                 & S (s1, s2, s3) &           &             &      & 3.469 &                            & $10.27$  & $1.73\pm0.01$ \\
GS\_4891         & C              & 53.07621  & $-27.86644$ & 1216 & 3.703 & 15872.7 \,|\, 3968.2       & $9.91$   & $1.61\pm0.01$ \\
                 & E              &           &             &      & 3.703 &                            & $8.64$   & $0.34\pm0.08$ \\
                 & N              &           &             &      & 3.703 &                            & $9.34$   & $0.73\pm0.05$ \\
COS\_HZ4         & C              & 149.61884 & $+2.05187$  & 1217 & 5.545 & 18206.9 \,|\, 3968.2       & $9.83$   & $1.73\pm0.01$ \\
COS\_HZ10        & E, C           & 150.24713 & $+1.55534$  & 1217 & 5.650 & 18206.9 \,|\, 3968.2       & $10.11$  & $1.83\pm0.02$ \\
                 & W              &           &             &      & 5.659 &                            & $10.10$  & $1.82\pm0.02$ \\
COS\_12306       & S, NW          & 150.12746 & $+2.32643$  & 1217 & 5.843 & 18206.9 \,|\, 3968.2       & $10.11$  & $2.21\pm0.01$ \\
HFLS3$^\dagger$  & S (s1, s2)     & 256.69917 & $+58.77320$ & 1264 & 6.359 & 7352.8 \,|\, 3559.7        & $-$      & $1.44\pm0.01$ \\
                 & W (w1, w2)     &           &             &      & 6.359 &                            & $-$      & $2.02\pm0.04$ \\
                 & g1             &           &             &      & 3.480 &                            & $-$      & $1.73\pm0.01$ \\
COS\_CR7         & A              & 150.24169 & $+1.80424$  & 1217 & 6.602 & 18206.9 \,|\, 3968.2       & $9.63$   & $1.70\pm0.01$ \\
                 & B              &           &             &      & 6.599 &                            & $9.10$   & $1.02\pm0.02$ \\
                 & C              &           &             &      & 6.598 &                            & $8.85$   & $0.79\pm0.03$ \\
COS\_3018        & C, NE, SE      & 150.12577 & $+2.26661$  & 1217 & 6.851 & 18206.9 \,|\, 3968.2       & $9.67$   & $1.84\pm0.01$ \\
SPT0311$-$58     & E, S, W, N1, N2              & 47.88858  & $-58.39264$ & 1264 & 6.919 &  7352.8 \,|\, 3559.7        & $10.39$  & $1.88\pm0.08$ \\
COS\_B14$-$65666 & C              & 150.41954 & $+1.91460$  & 1217 & 7.153 & 14705.6$^{\dagger\dagger}$ & $-$      & $1.63\pm0.08$ \\
                 & W              &           &             & 1217 & 7.149 &                            & $-$      & $1.96\pm0.11$ \\
                 & E              &           &             & 1217 & 7.153 &                            & $-$      & $1.46\pm0.15$ \\
MACSJ0416$-$Y1   & E, W           & 64.03926  & $-24.09319$ & 1208 & 8.312 & 18206.9 \,|\, 3968.2       & $9.61$   & $1.55\pm0.15$ \\
EGSY8P7          & C              & 215.03542 & $+52.89072$ & 1262 & 8.677 & 18206.9 \,|\, 3968.2       & $9.87$   & $1.99\pm0.14$ \\
MACS1149$-$JD1   & N, S           & 177.38992 & $+22.41269$ & 1262 & 9.110 & 18206.9 \,|\, 3968.2       & $8.76$   & $2.40\pm0.15$ \\
\bottomrule
\end{tabular}
\tablefoot{ID of the system, the galaxies/regions identified in each of them, the RA, DEC of the pointing, the JWST program, the redshift of the source and the exposure times. We also include the SFRs and stellar masses (M$_{\star}$) obtained for the large apertures highlighted in pink in Fig.~\ref{fig:integrated_oiii_maps} (see Sect.~\ref{sec:analysis}). Errors in the individual estimations log$_{10}$M$_{\star}$ are always smaller than 0.05~dex, although we consider an additional uncertainty of 0.15~dex to account for the systematics introduced by the different wavelength coverage in each source (see Sect.~\ref{subsec:stellar_mass_estimation}). $^{\dagger}$ \citet{jonesGANIFSJWSTNIRSpec2024} identified 4 galaxies at z~$\sim6.3$ (including `S' and `W') and a foreground galaxy (`g1') at $z\sim3.5$. $^{\dagger\dagger}$COS$\_$B14$-$65666 was only observed at medium spectral resolution (R1000). More details about each system are provided in Sect.~\ref{subsec:overview_systems}.}
\label{table:summary_observations}
\end{table*}

\begin{figure*}
\centering
\includegraphics[width=\textwidth]
{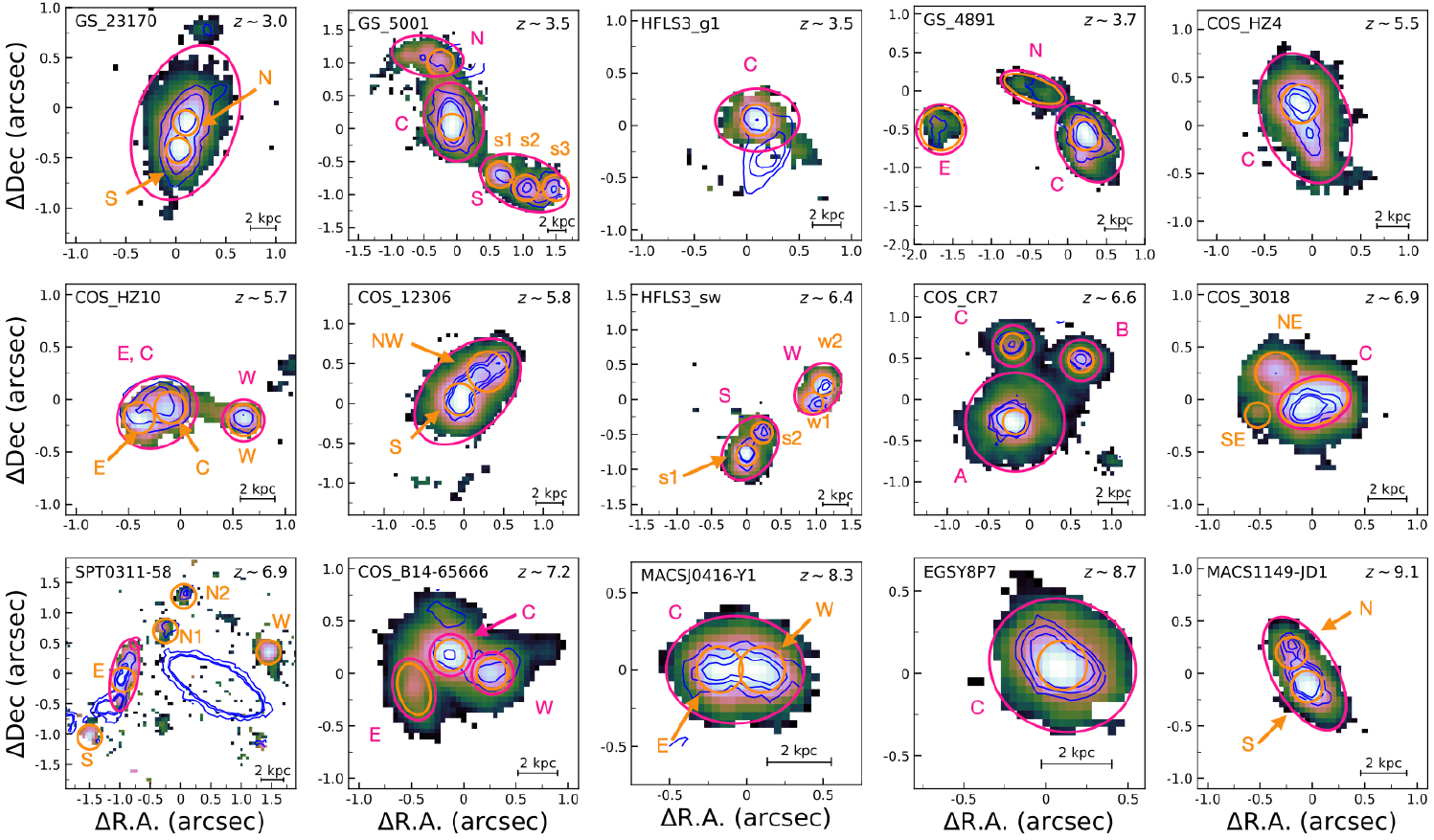}
\caption{Maps of integrated fluxes in \OIIIb for all the galaxy systems in our sample except HFLS3$\_$g1, for which we show integrated \Ha fluxes. Orange apertures mark small regions at the peaks of emission of the different (sub-) systems to constrain the dominant ionisation mechanisms (see Fig.~\ref{fig:integrated_line_diagnostics} and Sect.~\ref{subsec:diagnostics}). Pink apertures mark the larger regions used to study their extended emission. Blue contours correspond to the continuum emission at the 2500-3000\AA{} rest-frame at different levels, starting from three times the standard deviation of the background. Note that for HFLS3$\_$g1 at z=3.5, these contours also include the emission from the z=6.3 system. }
\label{fig:integrated_oiii_maps}
\end{figure*}

\begin{figure*}
\centering
\includegraphics[width=\textwidth]
{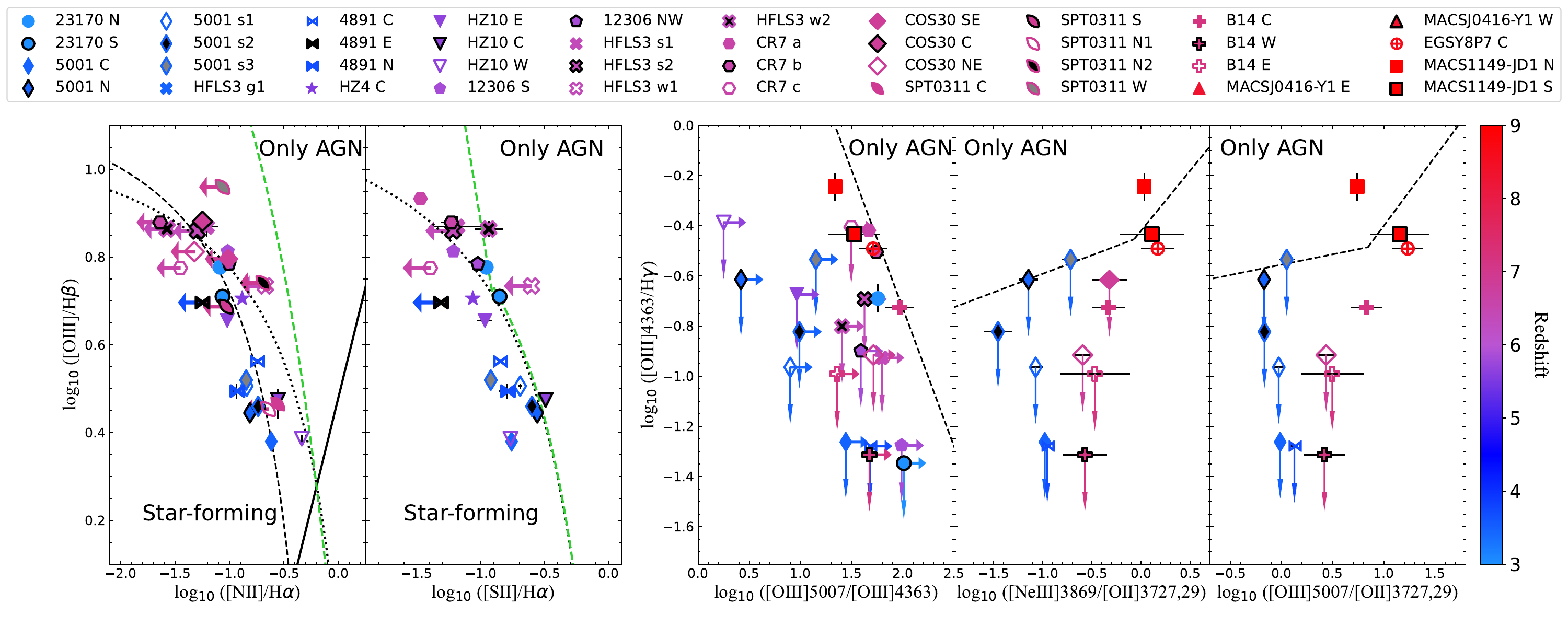}
\caption{Emission-line diagnostics for the galaxies in our sample obtained for the apertures highlighted in Fig.~\ref{fig:integrated_oiii_maps}. \emph{Left}: BPT-NII and BPT-SII diagrams \citep{baldwinClassificationParametersEmissionline1981a, veilleuxSpectralClassificationEmissionLine1987}. Dashed \citep{kauffmannHostGalaxiesActive2003b} and dotted black lines \citep{kewleyTheoreticalModelingStarburst2001b} demarcate the separation between ionisation coming from star formation and AGN activity based on studies of local galaxies. The dashed green lines in both plots have been recently proposed by \citet{scholtzJADESLargePopulation2025} to conservatively separate AGN galaxies (see Sect.~\ref{subsec:diagnostics} for details). \emph{Right}: Alternative diagnostic diagrams using \OIIIdm from \cite{mazzolariNewAGNDiagnostic2024}}.
\label{fig:integrated_line_diagnostics}
\end{figure*}

\subsection{Data reduction}
\label{subsec:data_reduction}

Raw data were homogeneously reduced with the {\it JWST} Science Calibration pipeline\footnote{\url{https://jwst-pipeline.readthedocs.io/en/stable/jwst/introduction.html}} version 1.15.0 under the Calibration Reference Data System context jwst\_1241.pmap. Several modifications to the default reduction were introduced to improve the data quality  \citep[see ][]{pernaGANIFSUltradenseInteracting2023}: count-rate frames were corrected for $1/f$ noise through a polynomial fit, regions affected by failed open MSA shutters and strong cosmic ray residuals were masked after calibration in Stage~2, while the remaining outliers were flagged in individual exposures using an algorithm similar to {\sc lacosmic} \citep{dokkumCosmicRayRejectionLaplacian2001}, rejecting the 95th (99.5th) percentile of the resulting distribution for the grating (R100/PRISM) observations. The reconstructed cubes of each dither position were combined using the `drizzle' method, providing a pixel scale of 0.05$\arcsec$ in the final datacube. Background subtraction was performed by generating a master background spectrum using spaxels away from the emitting sources. Finally, since the noise provided in the \texttt{ERROR} extension of NIRSpec/IFS datacubes can be underestimated compared to the actual noise in the data \citep[e.g.][]{rodriguezdelpinoGANIFSCoevolutionHighly2024}, the \texttt{ERROR} vectors were re-scaled based on the standard deviation in spectral regions free of emission lines.

\begin{figure}
\centering
\includegraphics[width=0.48\textwidth]
{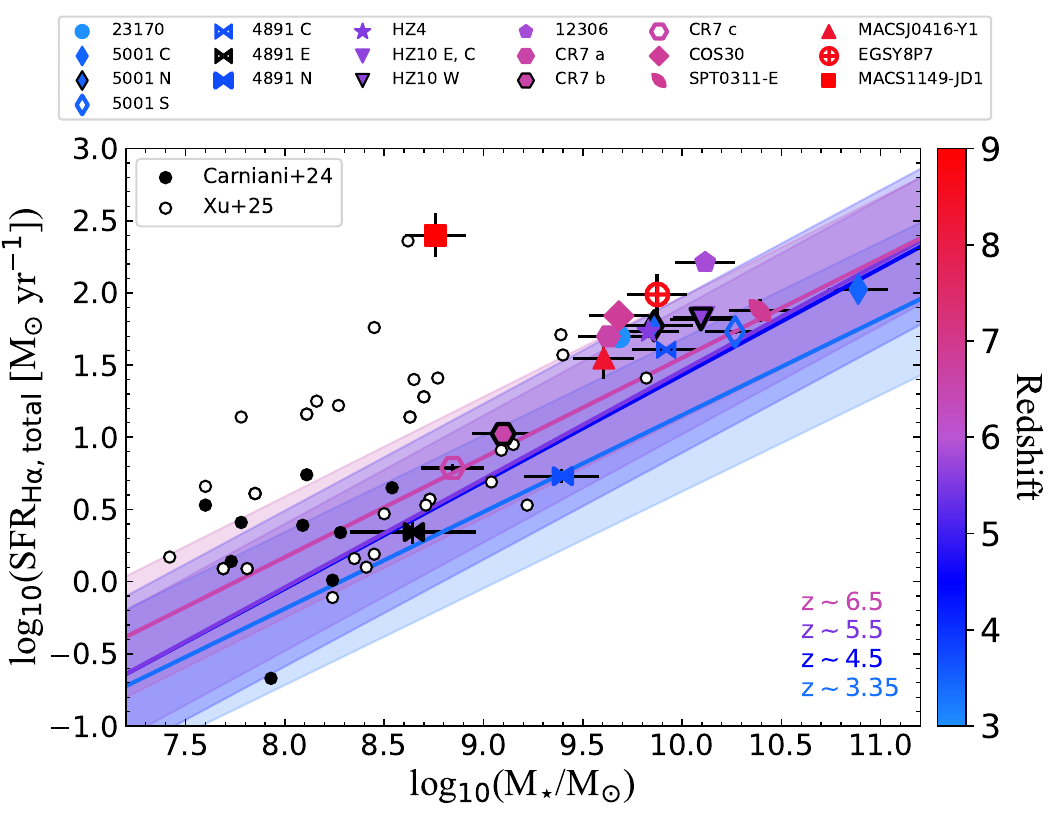}
\caption{Stellar mass vs total SFR for the individual star-forming galaxies in our sample colour-coded by the redshift of the source. The values displayed are estimated from the large apertures highlighted in pink in Fig.~\ref{fig:integrated_oiii_maps}. Solid lines and shaded regions correspond to the star formation main sequence at different redshift bins and the associated scatter from \citet{clarkeStarformingMainSequence2024a}. For comparison we show the sample of star-forming galaxies with outflows at $3<z<9$ from \citet{carnianiJADESIncidenceRate2024} and \citet{xuStellarAGNDrivenOutflows2025a}.}
\label{fig:stellar_mass_vs_sfr}
\end{figure}

\section{Analysis}
\label{sec:analysis}

\subsection{Spectral modelling of emission lines}
\label{subsec:spectral_modelling}

Through this study we constrained the emission-line fluxes and kinematics of the ionised gas through the spectral modelling of the R2700 spectra (R1000 in the case of the B14 system). This analysis is performed on a spaxel-by-spaxel basis and in the integrated spectra extracted from different regions of the galaxies in our sample. Firstly, we perform the fitting of individual, strong emission lines (i.e. \OIIIb, \Ha) using a single Gaussian function to map the structure of the ionised emission (see Fig.~\ref{fig:integrated_oiii_maps}). Secondly, a single Gaussian kinematic component is used to extract line fluxes from integrated regions by modelling the main optical emission lines available for each target (e.g. \OIIlines, \Hg, \OIIIdm, \Hb, \OIIIb, \NeIIIline, \Ha, \NIIb and \SIIlines; see Sect.~\ref{sec:general_properties}). Afterwards, we explore the presence of additional kinematic components in the spectra of our galaxies by including a second Gaussian kinematic component, with a larger velocity dispersion, in the spaxel-by-spaxel modelling of either \OIIIb or \Ha (see Sect.~\ref{sec:outflow_regions}). 
The emission line profiles are considered to favour a two-component model when: (1) the Bayesian information criterion (BIC) statistics \citep[][]{schwarzEstimatingDimensionModel1978} is significantly better ($\Delta$BIC$>10$) than for the one component
model and (2) the narrow and broad components present signal-to-noise ratio, $S/N$, $\geq3$ (measured as the maximum line flux over the scatter in the neighbouring continuum).

In the spectral modelling,\footnote{The kinematics maps and spectral modelling with one and two kinematic components of each source can be found at \url{https://doi.org/10.5281/zenodo.18032864}.} all emission lines associated with a kinematic component (in the one- and two-component cases) are fit together tying their kinematics. The ratios \OII$\lambda3729$/$\lambda3726$ and \SII$\lambda6716$/$\lambda6731$ are constrained within [0.3839, 1.4558] and [0.4375, 1.4484], respectively, corresponding to the theoretical limits for low (1~cm$^{-3}$) and high (10$^5$~cm$^{-3}$) density regimes \citep{sandersMOSDEFSurveyElectron2016}. We assume the \citet{calzettiDustContentOpacity2000} attenuation law and intrinsic values \Ha/\Hb~=~2.860 and \Hg/\Hb~=~0.468, corresponding to an electron temperature $T_{\rm e}=10^{4}$~K and electron density of $n_{\rm e}=100$~cm$^{-3}$ for case B recombination \citep{osterbrockAstrophysicsGaseousNebulae2006}. The spectral modelling is performed using a Bayesian approach based on Markov Chain Monte Carlo (MCMC) techniques. In particular, we employ the \textsc{EMCEE} software developed by \citet{foreman-mackeyEmceeMCMCHammer2013b}. The reported 1$\sigma$ uncertainties on each parameter are calculated as half the difference between the 16th and 84th percentiles of the posterior probability density distribution. The spectral modelling takes into account the wavelength-dependent spectral resolution of NIRSpec \citep{jakobsenNearInfraredSpectrographNIRSpec2022a}.

\subsection{Stellar mass estimation}
\label{subsec:stellar_mass_estimation}
The available R100 NIRSpec/IFS observations for all galaxies in our sample provide access to their stellar emission, allowing the estimation of stellar masses. The only exceptions are the B14 system, which was observed in the R1000 configuration and lacks the depth to detect the stellar emission, and HFLS3, where the continuum emission is very weak. For the computation of the stellar masses we employ the integrated spectra extracted from the large apertures that encompass the extended emission of the ionised gas (pink apertures in Fig.~\ref{fig:integrated_oiii_maps}). The modelling of the spectral energy distribution is performed with the software \verb|Bagpipes| \citep{carnallHowMeasureGalaxy2019}, that uses the \citet{bruzualStellarPopulationSynthesis2003a} stellar population models. We assume the same extinction for the stellar and nebular component and a \citet{calzettiDustContentOpacity2000} dust attenuation law, with $A_{\rm V}$ varying between 0 and 3.0. The values for the ionisation parameter (log$U$) and the metallicity ($Z$) are allowed to vary within the ranges [$-3.5, -2.0$] and [$0.1-1$~Z$_{\odot}$], respectively, based on the values measured from their nebular emission \citep[e.g.][]{lampertiGANIFSJWSTNIRSpec2024a, rodriguezdelpinoGANIFSCoevolutionHighly2024}. We consider a  delayed exponential star formation history with a recent burst. For SPT0311-58-E, MACSJ0416-Y1 and MACS1149-JD1, which are lensed systems, stellar masses are estimated after correcting for magnification factors, $\mu$, of 1.15 \citep{arribasGANIFSCoreExtremely2024a}, 1.43 \citep{kawamataPRECISESTRONGLENSING2016} and 10  \citep{marconciniGANIFSInterplayMerger2024a}, respectively. Since our sample spans a wide range in redshift, the changing coverage of the rest-frame emission in the R100 data could have an impact in the estimation of the stellar masses. To evaluate this effect, we have repeated the estimation of stellar masses constraining the data to a common rest-frame wavelength range, namely $1500\AA$–$4945\AA$. In this test we find an average offset of only $\sim$~$0.15$~dex towards lower stellar mass values and no variation of this offset with redshift. Thus, since the impact in our results is negligible, we maintain the stellar mass values estimated with the full wavelength coverage of the R100 data but add a 0.15~dex to the errors in their estimation to account for these systematic uncertainties. The estimated stellar masses for the galaxies/regions in our sample range within log$_{10}$~(M$_\star$/M$_\odot$)~=~$8.5-11$. The individual values are included in Table~\ref{table:summary_observations}.

\section{General properties of the sample}
\label{sec:general_properties}
We started the characterisation of the sample of star-forming systems by exploring the distribution of the ionised gas. We traced this gas through the \OIIIb emission line, which is observed for all systems in our sample with the exception of HFLS3$\_$g1 (the additional galaxy at $z\sim3.5$ discovered in the foreground of the HFLS3 system at $z=6.3$ by \citealt{jonesGANIFSJWSTNIRSpec2024}), for which we use \Ha. The maps of integrated line fluxes (obtained as described in Sect.~\ref{subsec:spectral_modelling} and when $S/N>3$), without applying a correction for dust attenuation, are shown in Fig.~\ref{fig:integrated_oiii_maps}, with blue contours highlighting the continuum emission at 2500-3000\AA{} rest-frame. These maps of the ionised gas distribution illustrate the complexity of the systems, in many cases composed by several galaxies/regions. 

Within these 15 complex systems, we identified numerous regions characterised by strong \OIIIb and/or continuum emission. Based on this emission, we define a total of 40 regions that are associated with individual, detached galaxies, and to the peaks of \OIIIb emission. These regions are highlighted in orange in Fig.~\ref{fig:integrated_oiii_maps} and listed in Table \ref{table:summary_observations}. We also define larger apertures encompassing the more extended emission, which are displayed in pink. Although we did not detect continuum emission in the regions defined in COS\_3018 (hereafter COS30) and B14, we note that such emission has been confirmed in NIRCam images presented in \citet{scholtzGANIFSISMProperties2025} and \citet{jonesGANIFSInterstellarMedium2026}. For the subsequent analysis, we extracted the integrated R100 and R2700 spectra within these apertures, applying aperture corrections as described in Appendix~\ref{app:aperture_corrections}.

\subsection{Star formation as a dominant ionisation mechanism}
\label{subsec:diagnostics}
Since for this study we are interested in the characterisation of feedback from star formation activity we need to work with a sample of galaxies that present no clear evidence of the presence of an AGN. To confirm the star-forming nature of these complex systems and rule out clear evidence of AGN activity we study the spectra of the individual, nuclear regions (defined as orange apertures in Fig.~\ref{fig:integrated_oiii_maps}) that dominate the emission from the ionised gas. To determine the dominant ionisation mechanism in each region, we explore in Fig.~\ref{fig:integrated_line_diagnostics} the location of the measured integrated line flux ratios (see Sect.~\ref{subsec:spectral_modelling}) in the standard BPT-[NII] and BPT-[SII] diagnostic diagrams \citep{baldwinClassificationParametersEmissionline1981a, veilleuxSpectralClassificationEmissionLine1987} as well as in the diagnostic diagrams recently developed by \citet{mazzolariNewAGNDiagnostic2024} which are based on \OIIIdm. The BPT diagrams include reference lines from studies of local galaxies that demarcate the regimes that could be explained considering only ionisation from star formation (below the black, dashed line from \citealt{kauffmannHostGalaxiesActive2003b}) and the one where theoretical models require the presence of AGN activity to explain the ionisation state of the gas (black, dotted lines from \citealt{kewleyTheoreticalModelingStarburst2001b}). In addition, we have also added the demarcation line (green) recently proposed by \citet{scholtzJADESLargePopulation2025} to conservatively separate AGNs. We only include measurements where the peak of the emission line has $S/N>3$ and employ $1\sigma$ upper limits for the lines without a clear detection. The systems/galaxies included in each panel varies because of the different wavelength coverage of their rest-frame emission. The use of the BPT-[NII] diagram in high-redshift studies has raised concern because low metallicity AGNs, typical of early cosmic epochs, tend to intrinsically present lower \NII/\Ha ratio, thus shifting towards the region of the diagram associated with ionisation from star formation \citep{feltreNuclearActivityStar2016, ublerGANIFSMassiveBlack2023c, scholtzJADESLargePopulation2025}. This effect is not relevant for \SIIlines, although these lines are generally weak and/or outside the NIRSpec wavelength coverage at $z>6.5$. Alternatively, the diagrams from \citet{mazzolariNewAGNDiagnostic2024} provide a cleaner selection of AGNs as their higher ionisation leads to a boost in the \OIIIdm emission compared to star-forming galaxies. However, these diagrams only allow for the identification of clear AGN cases and cannot be used to discard the presence of AGNs in systems without strong \OIIIdm emission. Despite these inherent caveats, we note that most of our sources lie in the regions of the diagrams that are not dominated by ionisation from AGNs. In the BPT-[NII] diagram all sources are on the left of the demarcation line by \citet{scholtzJADESLargePopulation2025} while, in the BPT-[SII], three sources overlap with this conservative demarcation line, and only one (HFLS3$\_$w1) presents upper limits in the \SIIlines that are compatible with a clear AGN classification. A similar result is found in the \OIIIdm-based diagrams (three left panels in Fig.~\ref{fig:integrated_line_diagnostics}) where only MACS1149-JD1-N is located in the "Only AGN" region, just above the demarcation line, in two of the diagrams. Despite the possible presence of an AGN in this galaxy, we keep it in our sample, bearing in mind that any outflow detected might also be driven by AGN activity. In the case of HFLS3$\_$w1, given that its possible AGN nature is only suggested by the BPT-[SII] using upper limits, we consider it as a star-forming system not dominated by AGN activity. Based on these diagnostic diagrams, and considering both the absence of X-ray emission and the lack of AGN-associated broad-line regions, we consider the classification of all other galaxies in the sample as star-forming galaxies to be robust. Finally, as shown in the last two panels of Fig.~\ref{fig:integrated_line_diagnostics}, the \NeIIIline/\OIIlines and \OIIIb/\OIIlines ratios clearly increase at higher redshifts, which is indicative of a higher ionisation parameter at earlier epochs, as identified in other works \citep[e.g.][]{cameronJADESProbingInterstellar2023a}.



\subsection{Comparison with the general population of star-forming galaxies}
\label{subsec:main_sequence}
In this section we compare the SFRs and stellar masses of our individual systems with the general population of star-forming galaxies at similar redshifts. We derive the SFRs using the line fluxes in \Ha and \Hb measured in the one-component fit to the integrated spectra extracted from the 
same apertures used for the stellar mass estimates. We note that, although many of these systems present ionised outflows, most of the line flux extracted from the large apertures is recovered with a single component fit. The estimated line fluxes (see Sect.~\ref{subsec:spectral_modelling}) are corrected for aperture losses and for magnification when required. We determine the nebular dust attenuation in our galaxies by comparing the observed and intrinsic ratios of a pair of Balmer emission lines, as in \citet{dominguezDustExtinctionBalmer2013}. At $z<7$ we work with \Ha and \Hb, while for higher redshift galaxies we employ \Hb and \Hg. In the case of HFLS3$\_$g1 we do not cover any other hydrogen emission line apart from \Ha and no dust correction can be estimated; thus, we compute lower limits on SFR. Finally, to convert \Ha luminosities to SFR we apply the calibration from \citet[][; Eq. 1]{clarkeStarformingMainSequence2024a}, considering stellar metallicities $Z_{*}=0.004$, which correspond to 0.29~Z$_\odot$ (12 + log(O/H)~=~8.15), and assume a \citet{chabrierGalacticStellarSubstellar2003} IMF. The adopted metallicity for this calibration is consistent with the values of 12 + log(O/H) measured in our work (see Sect.~\ref{subsec:results_av_z}). For the galaxies at $z>7$ where we do not cover \Ha, we infer its SFR from the \Hb and \Hg and the estimated dust attenuation.  

In Fig.~\ref{fig:stellar_mass_vs_sfr} we show the total SFR as a function of the stellar mass for our sample of star-forming galaxies. As a reference, we include the `star formation main sequence' relations derived at different redshift bins ([2.7,4.0], [4.0,5.0], [5.0, 6.0], [6.0,7.0]) by \citet{clarkeStarformingMainSequence2024a}, representing the general population of star-forming galaxies. Since we do not provide a star formation main sequence beyond $z=7$, more distant galaxies (redder points in the plot) do not have a reference to compare with. We also add the samples of galaxies with ionised outflows at $3<z<9$ from \citet{carnianiJADESIncidenceRate2024} and \citet{xuStellarAGNDrivenOutflows2025a}, which also assume a \citet{chabrierGalacticStellarSubstellar2003}  IMF. In general, our galaxies lie above the star formation main sequence at their corresponding redshifts, indicating that most of them are observed in a star-bursting phase. Samples from previous works, especially in \citet{xuStellarAGNDrivenOutflows2025a}, also present SFRs higher than those expected given their stellar masses. Besides that, our sample encompasses the higher mass end of the galaxy populations, with most of the sample having stellar masses within log$_{10}$~(M$_\star$/M$_\odot$)~=~$8.5-11$, while previous works generally contain galaxies with log$_{10}$~(M$_\star$/M$_\odot$)$<9.5$. 
 
\begin{figure*}
\centering
\includegraphics[width=\textwidth]
{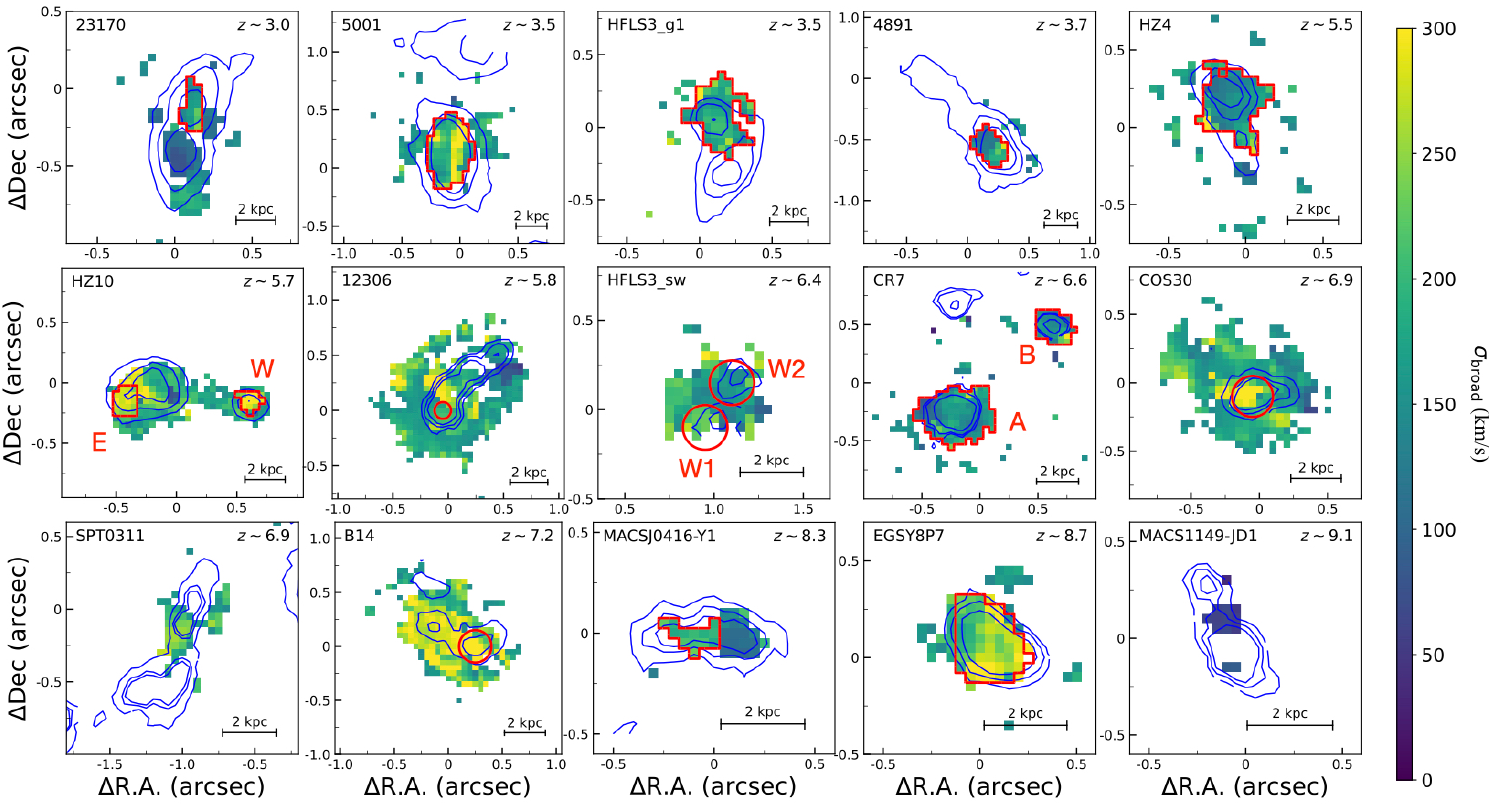}
\caption{Velocity dispersion maps of the broad component ($\sigma_{\rm broad}$) for all galaxies in our sample. The regions identified as outflows are highlighted in red. No outflows are identified in SPT0311-58 and MACSJ1149-JD1. Blue contours correspond to the stellar continuum emission, as in Fig.~\ref{fig:integrated_oiii_maps}. The properties we derived for regions hosting outflows are listed in Table~\ref{table:outflow_properties}.}
\label{fig:sigma_broad_maps}
\end{figure*}

\section{Identification of the regions hosting outflows}
\label{sec:outflow_regions}

The identification of galactic outflows in complex systems such as those in our sample, potentially undergoing mergers and tidal interactions, is challenging because these processes enhance turbulence, produce tidal structures, and create overlapping kinematic components that can mimic the spectral broadening typically associated with outflows. To minimise these uncertainties, we rely on the spatially resolved information of the gas kinematics and the stellar continuum emission provided by the NIRSpec IFS R2700 and R100 data. Using this information, we define regions potentially hosting ionised outflows as those presenting a secondary broad kinematic component with high velocity dispersion ($\sigma_{\rm broad}>100$~km/s) or a single broad kinematic component with even higher velocity dispersion ($>140$~km/s), as in \citet{lampertiGANIFSJWSTNIRSpec2024a}. We also ensure that these regions overlap with or are connected to the stellar continuum emission, as expected if the outflows are driven by star formation. By applying this constrain we exclude detached gaseous regions that may be associated with tidally induced turbulence, where we expect very low (i.e. undetectable) continuum emission at the high redshifts probed in our study. This approach, however, may result in missing some outflows that, due to projection effects, are not co-spatial with the stellar emission.

In Fig.~\ref{fig:sigma_broad_maps} we present the velocity dispersion maps of the broad component together with the continuum emission at 2500-3000\AA{} rest-frame (blue contours). The regions that are identified as potentially hosting ionised outflows are highlighted with red contours. Additionally, in Appendix~\ref{app:kinematic_maps} we show the individual maps of emission-line fluxes, velocity and velocity dispersion for the one- and two-component fits.

Even though we acknowledge that the definition of the regions hosting outflows might be uncertain for some objects due to their complex nature, our IFS data yield much better information than previous works based on MOS and WFSS spectra \citep[e.g.][]{carnianiJADESIncidenceRate2024, xuStellarAGNDrivenOutflows2025a} because they allowed us to not only determine the extent of the broad component in a better way but also isolate the emission from the region potentially hosting the outflows and therefore adequately constrain the properties of the outflowing gas. Moreover, as we show later in the paper, our results suggest that the present identification of the broad component regions with outflows is globally correct, since the metallicity in the broad component is, on average, enhanced with respect to the hosts (see Sect.~\ref{subsec:results_av_z} and ~\ref{subsec:discussion_metals}). This behaviour is not expected if the broad component is the result of a mixing mechanism, such as those associated with  interactions and mergers. However, for some individual sources the definition of regions hosting outflows on the basis of these admittedly simple criteria may be dubious. For instance, we find some cases with large red-shifted velocities ($\Delta$$V\leq-200$~kms$^{-1}$) with respect to the host galaxy (HZ10$\_$E and B14) that challenge the outflow scenario, since the redshifted component is expected to be obscured by the host galaxy. Even though these high-z galaxies have relatively low extinction (see Sect.~\ref{subsec:results_av_z} and Table~\ref{table:outflow_properties}) and for some geometries the receding gas might appear to be in front of the galaxy from the observer's point of view, we consider these broad components as `candidate' outflows (see further discussion in Sect.~\ref{subsec:discussion_metals}). Thus, although the measured properties of these two regions are listed in Table~\ref{table:outflow_properties}, we do not include them in our global analysis of the properties of ionised outflows in our sample. Finally, only two systems, SPT0311-58 and MACS1149-JD1, do not present evidence of hosting ionised outflows. In some galaxies (4891, 5001, HFLS3$\_$g1, HZ4, CR7$\_$a, CR7$\_$b, EGSY8P7) we consider most of the extension of the broad emission as the region hosting the outflow, since they display relatively smooth variations in the velocity dispersion and largely overlap with the stellar continuum emission. In other galaxies that display a clumpier structure (23170, HZ10$\_$E, HZ10$\_$W, MACSJ0416-Y1), we only consider the regions in the galaxy with the highest velocity dispersion. In the most complex systems that are mergers (12306, HFLS3$\_$w (1, 2), COS30 and B14) we adopt a conservative approach by considering only the nuclear region around the peak of stellar continuum emission, where the gas is expected to have shorter relaxation times and virialise faster, thus making a broad component more likely to be associated with outflowing gas. In total, after applying our outflow identification criteria and removing potential cases of mergers causing the broad component emission, we end up with a total of 16 regions hosting ionised outflows.

To study the properties of the outflowing gas we extract the integrated spectra from the regions identified above (applying aperture corrections as described in Appendix~\ref{app:aperture_corrections}) and perform a two-component modelling of the main emission lines, tying the kinematics of all lines in the narrow and broad kinematic components. In this spectral modelling we apply the same constrains described in Sect.~\ref{subsec:spectral_modelling}. We evaluate whether other emission lines apart from \OIIIb and \Ha present a broad kinematic component by applying the criteria described in Sect.~\ref{sec:outflow_regions} individually for each line, finding that a broad component is frequently detected in \Hg, \Hb, \Ha and \NII. We show in Appendix~\ref{app:outflow_fits} the modelling of the integrated outflow region in each galaxy, which includes the best fit model (one or two component) for each emission line.

\section{Results}
\label{sec:results}

\subsection{Overview of individual systems}
\label{subsec:overview_systems}

As commented above, the presence of significant substructures and/or mergers in the systems in our sample poses a challenge in the identification of the regions that may host outflows since they also lead to turbulence, tidal tails and overlap between different galaxies that might also produce a spectral broadening similar to that characteristic of outflows. For this reason, the definition of the outflowing regions requires a detailed inspection of the global kinematics of the gas, along with its connection to the stellar continuum emission coming from starburst region leading to the star formation--driven outflow. In Appendix~\ref{app:individual_systems} we analyse each galaxy/system in detail and define the regions considered to host ionised outflows. We base this definition on the velocity dispersion maps presented in Fig.~\ref{fig:sigma_broad_maps} and the individual kinematic maps also presented for each galaxy in Appendix~\ref{app:individual_systems}. 

We note here that the properties derived in this work (dust attenuation, oxygen abundance, kinematics) are consistent with those obtained in previous works on individual galaxies \citep[e.g.][]{lampertiGANIFSJWSTNIRSpec2024a, rodriguezdelpinoGANIFSCoevolutionHighly2024, parlantiGANIFSMultiphaseAnalysis2025}, but the actual values we obtained might differ slightly because of the different data processing steps (e.g. aperture corrections), selected regions and outflow selection criteria. A description of the findings in each system is given in Appendix~\ref{app:individual_systems}.

\subsection{Characterisation of the outflowing gas}
\label{subsec:characterisation_outflows}
In this section we constrain the main properties of the ionised outflows identified in our sample. We determine the spatial extent of the outflows, use the integrated spectra of these regions to estimate the dust attenuation and gas metallicity of the outflowing gas and compare them to those of the host galaxies, and compute the outflowing gas kinematics, exploring possible correlations with the properties of the host regions. 

\subsubsection{Determination of the outflow size}
\label{subsubsec:size_determination}

The determination of the outflow sizes in the literature has been performed using a variety of methods. In recent studies of ionised outflows in high-redshift star-forming galaxies that used NIRSpec Multi-object spectroscopy and NIRCam Wide Field Slitless Spectroscopy \citep{carnianiJADESIncidenceRate2024, cooperHighvelocityOutflowsOIII2025, xuStellarAGNDrivenOutflows2025a}, the extent of the outflow was considered to be the same as the host galaxy, thus employed the half-light radius of the source estimated from continuum imaging as the radius of the outflow. Our resolved velocity dispersion maps demonstrate that this approach might be valid for some sources where the outflow has a similar extension as the continuum emission (e.g. HZ4, EGSY8P7) but, in other cases, it can lead to underestimation (CR7$\_$a) or overestimation (4891) of the true extent of the outflow (see Fig.~\ref{fig:sigma_broad_maps}). These mismatches are likely related to the morphology of the source (clumpy or disk-like), the merging state and the orientation of the system with respect to the observer. In their study of \CII outflows in main-sequence galaxies at $z\sim5$, \citet{birkinALMACRISTALSurveyWeak2025} assume an outflow radius of 6~kpc, following the sizes estimated from the stacking of marginally spatially resolved \CII emission at high velocities by \citet{ginolfiALPINEALMAIISurvey2020}. 

\renewcommand{\arraystretch}{1.2} 
\begin{table*}
\caption{Integrated properties of the ionised outflows identified in each galaxy or region (red contours in Fig.~\ref{fig:sigma_broad_maps}).}
\centering
\small 
\begin{tabular}{lcccccccccc}
\toprule
\makecell{Outflow host \\ (Galaxy or region)} & 
\makecell{V$_{\mathrm{out}}$ \\ (\si{\kilo\meter\per\second})} & 
\makecell{$\mathrm{\sigma}_{\mathrm{out}}$ \\ (\si{\kilo\meter\per\second})} & 
\makecell{$A_{V,\mathrm{narrow}}$} & 
\makecell{$A_{V,\mathrm{broad}}$} & 
\makecell{$Z_{\mathrm{narrow}}$} & 
\makecell{$Z_{\mathrm{broad}}$} & 
\makecell{$r_{\mathrm{out}}$ \\ (kpc)} &
\makecell{log$_{10}SFR_{\mathrm{out}}$ \\ (M$_\odot$yr$^{-1}$)} & 
\makecell{$\eta=\dot{M}_{\rm out}/SFR_{\mathrm{out}}$} \\
\midrule
GS$\_$23170      & 179 $\pm$ 8   & 135.7 $\pm$ 4.4 & 0.05$_{-0.05}^{+0.09}$ & 1.57 $\pm$ 0.12 & 8.26 $\pm$ 0.07             & $-$ & 0.9 & 1.5 $\pm$ 0.4 & 1.5 $\pm$ 0.37 \\
GS$\_$5001       & 218 $\pm$ 6   & 172.5 $\pm$ 3.4 & 0.77 $\pm$ 0.04 & 2.54 $\pm$ 0.09 & 8.52 $\pm$ 0.04 & 8.51 $\pm$ 0.05  & 1.9 & 2.0 $\pm$ 0.7 & 0.13 $\pm$ 0.02 \\
GS$\_$4891            & 272 $\pm$ 9   & 157.6 $\pm$ 4.6 & 1.39 $\pm$ 0.06 & 0.18 $\pm$ 0.14 & 8.33 $\pm$ 0.06 & 8.57 $\pm$ 0.06  & 1.1 & 1.5 $\pm$ 0.1 & 0.01 $\pm<0.01$ \\
COS$\_$HZ4            & 199 $\pm$ 3   & 145.5 $\pm$ 1.9 & 0.27 $\pm$ 0.07 & 0.95 $\pm$ 0.11 & 8.25 $\pm$ 0.06 & 8.41 $\pm$ 0.06  & 1.4 & 1.6 $\pm$ 0.3 & 0.39 $\pm$ 0.08 \\
COS$\_$HZ10$\_$E$^{*}$   & 480 $\pm$ 12  & 259.8 $\pm$ 6.2 & 0.80 $\pm$ 0.14 & 1.75 $\pm$ 0.18 & 7.59 $\pm$ 0.29 & 8.47 $\pm$ 0.07  & 0.7 & 1.2 $\pm$ 0.3 & 0.55 $\pm$ 0.14 \\
COS$\_$HZ10$\_$W  & 416 $\pm$ 26  & 264.0 $\pm$ 13.0 & 1.57 $\pm$ 0.36 & 2.41 $\pm$ 0.21 & 8.55 $\pm$ 0.08 & 8.56 $\pm$ 0.06 & 0.6 & 1.7 $\pm$ 0.8 & 1.72 $\pm$ 0.65 \\
COS$\_$12306            & 313 $\pm$ 8  & 175.2 $\pm$ 4.2 & 1.26 $\pm$ 0.18 & 1.01 $\pm$ 0.07 & 8.2 $\pm$ 0.09 & 8.19 $\pm$ 0.08 & 1.7 & 1.7 $\pm$ 0.5 & 1.0 $\pm$ 0.23 \\
HFLS3$\_$w1           & 297 $\pm$ 16  & 230.4 $\pm$ 8.3 & 0.67 $\pm$ 0.43 & 2.66 $\pm$ 0.31 & 8.23 $\pm$ 0.14 & 8.39 $\pm$ 0.09  & 0.7 & 1.8 $\pm$ 1.1 & 5.06 $\pm$ 2.59 \\
HFLS3$\_$w2           & 317 $\pm$ 11  & 156.3 $\pm$ 5.5 & 1.84 $\pm$ 0.23 & 1.18 $\pm$ 0.26 & 7.99 $\pm$ 0.18 & 8.28 $\pm$ 0.14  & 0.7 & 1.6 $\pm$ 0.8 & 0.36 $\pm$ 0.15 \\
HFLS3$\_$g1           & 462 $\pm$ 88  & 330.7 $\pm$ 44.2 & $-$            & $-$            & $-$            & $-$                & 1.7 & > 0.6 & $<0.01$ \\
COS$\_$CR7$\_$a       & 189 $\pm$ 2   & 153.9 $\pm$ 1.2 & 0.01$_{-0.01}^{+0.03}$ & 0.88 $\pm$ 0.09 & 7.83 $\pm$ 0.14 & $-$       & 1.6 & 1.7 $\pm$ 0.3 & 0.77 $\pm$ 0.3 \\
COS$\_$CR7$\_$b       & 169 $\pm$ 4   & 142.4 $\pm$ 2.3 & 0.01$_{-0.01}^{+0.06}$ & 0.48 $\pm$ 0.21 & 7.96 $\pm$ 0.18 & $-$       & 0.9 & 1.0 $\pm$ -0.1 & 0.8 $\pm$ 0.39 \\
COS$\_$3018           & 448 $\pm$ 18  & 334.3 $\pm$ 9.0 & 1.00 $\pm$ 0.04$^{\dagger}$ & $-$ & 7.94 $\pm$ 0.11 & $-$              & 0.8 & $>1.2$        & 0.09 $\pm <0.01$ \\
COS\_B14$-$65666$^{*}$            & 586 $\pm$ 10  & 324.1 $\pm$ 5.3 & 1.73 $\pm$ 0.32$^{\dagger}$   & $-$ & 8.11 $\pm$ 0.22 & 8.28 $\pm$ 0.41  & 0.8 & 1.6 $\pm$ 1.0 & 0.58 $\pm <0.01$ \\
MACSJ0416-Y1          & 251 $\pm$ 14  & 193.7 $\pm$ 7.5 & 0.18$_{-0.18}^{+0.26}$$^{\dagger}$ & $-$ & 7.90 $\pm$ 0.22 & $-$              & 0.5 & $>1.2$         & 1.55 $\pm$ 1.05 \\
EGSY8P7               & 266 $\pm$ 5   & 214.3 $\pm$ 2.6 & 0.74 $\pm$ 0.35 & 1.27 $\pm$ 0.68 & 7.67 $\pm$ 0.23 & $-$              & 1.0 & 2.3 $\pm$ 1.9 & 0.66 $\pm$ 0.61 \\
\bottomrule
\end{tabular}
\tablefoot{The terms $V_{\rm out}$ and $\sigma_{\rm out}$ are the maximum velocity and velocity dispersion of the outflows, whereas $A_{V,\mathrm{narrow}}$ and $A_{V,\mathrm{broad}}$ are the dust attenuation in the narrow (systemic) and broad (outflow) kinematic components, estimated from \Ha and \Hb fluxes (at $z<7$) and from \Hb and \Hg fluxes (at $z>7$). $A_{V,\mathrm{broad}}$ is only provided when a broad component is clearly detected in both emission lines. The terms $Z_{\mathrm{narrow}}$ and $Z_{\mathrm{broad}}$ are the oxygen abundances (12+log(O/H)) in the narrow and broad components, respectively. We provide $Z_{\mathrm{broad}}$ values for galaxies in which we detect broad emission lines in \Hb, \OIII, \Ha, and \NII, and we were able to estimate the metallicity using the R3 and N2 indicators. The term $r_{\rm out}$ is the size of the outflow, SFR$_{\rm out}$ is the SFR within the region hosting the outflow (estimated using the flux from the narrow and broad components; see Sect.~\ref{subsec:constraining_kinematics}), and $\eta$ is the mass-loading factor. $^{*}$Candidate outflows. $^{\dagger}$$A_{V,\mathrm{narrow}}$ values estimated from the single component fit. }
\label{table:outflow_properties}
\end{table*}
\renewcommand{\arraystretch}{1} 

In our case, we benefit from the spatially resolved information provided by NIRSpec IFS data that allows the spatial extent of the broad emission to be constrained. Other works on ionised outflows using NIRSpec IFS data have estimated the radius of the outflow as the distance from the centre of the galaxy to the region with high velocity dispersion \citep[e.g.][]{lampertiGANIFSJWSTNIRSpec2024a} or have considered a flux-weighted approach \citep{bertolaGANIFSMapping352025}. In our sample, 
as can be seen in Fig.~\ref{fig:sigma_broad_maps}, the morphology and size of the regions hosting the outflows is very heterogeneous. We follow two approaches to define the size of the regions hosting outflows. On the one hand, in the cases where the region with high velocity dispersion is relatively isolated and associated with continuum emission (23170, 5001, HFLS3$\_$g1, 4891, HZ4, HZ10 (E, W), CR7 (a,b), MACSJ0416-Y1, EGSY8P7), we computed the radius of the outflow, $r_{\rm out}$, as the circularised radius of the region hosting the outflow (i.e. the area of the regions highlighted in red in Fig.~\ref{fig:sigma_broad_maps}), as in \citet{rodriguezdelpinoGANIFSCoevolutionHighly2024}. On the other hand, in the cases where the broad emission presents a large extension beyond the continuum emission, suggesting that tidal interactions or mergers could also be inducing the broad emission (HFLS3$\_$w (1,2), 12306, B14, COS30), we defined small circular regions with $r_{\rm out}=0.25-0.3\arcsec$ around the peak of continuum emission, assuming that in these cases the outflow could have a larger extension. We also note that our size estimations should be considered as lower limits, since what we observe are projected sizes on the sky. The estimated values of $r_{\rm out}$ are listed in Table~\ref{table:outflow_properties}. The ionised outflows in our sample have a mean extent of $r_{\rm out}$=1.1~kpc, ranging from a minimum of 0.5~kpc (MACSJ0416-Y1) to a maximum of 1.7~kpc (HFLS3$\_$g1, 12306). These sizes are generally larger than those estimated by \citet{carnianiJADESIncidenceRate2024} for their sample of outflows in lower-mass (log$_{10}$~(M$_\star$/M$_\odot$)$<9.5$) galaxies at $3<z<9$ based on the continuum emission from the host, finding in most of the cases $r_{\rm out}<1$~kpc and a mean size $r_{\rm out}=0.67$~kpc.

\begin{figure*}
\sidecaption
    \includegraphics[width=0.7\textwidth]
{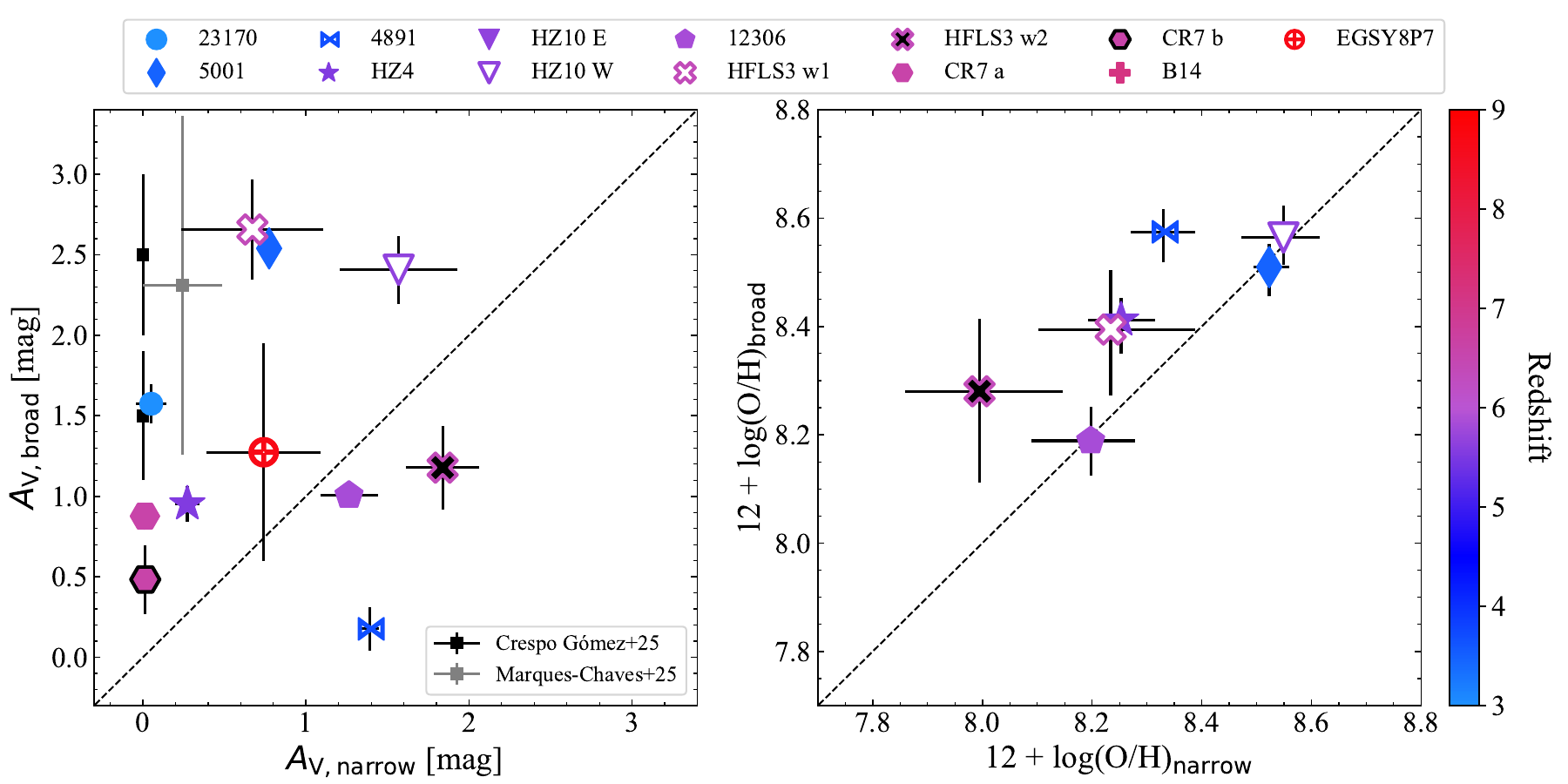}
    \caption{Comparison of the nebular dust attenuation (\emph{left}) and metallicity (\emph{right}) for the narrow and broad kinematic components in the galaxy regions in our sample identified to host ionised outflows. Symbols are colour-coded according to the redshift of the galaxies. We include values measured in other star-forming galaxies at $z\sim6.1-6.2$ by \citet{crespo-gomezRIOJADustyOutflows2025} and \citet{marques-chavesExtremelyUVbrightStarbursts2025}. The black dashed lines correspond to the one-to-one relation.}
\label{fig:extinction_metallicity}
\end{figure*}

\subsubsection{Dust attenuation and gas metallicity of the outflows}
\label{subsec:results_av_z}
The detection of a broad kinematic component across multiple emission lines in the integrated spectra of the outflows offers the opportunity to probe the properties of the outflowing gas in detail and directly compare them with those of the host galaxy. In this section we focus on the study of the dust attenuation and metallicity of the different gas components. 

Applying the same method described in Sect.~\ref{subsec:main_sequence}, we determine the dust attenuation in the narrow (systemic) and broad (outflow) kinematic components using a pair of Balmer emission lines, \Ha and \Hb for galaxies at $z<7$ and \Hb and \Hg for those at $z>7$. We could perform an estimation of $A_{V,\mathrm{broad}}$ only when a broad component is clearly detected ($S/N>3$) in both emission lines. This estimation is possible in the integrated spectra of 12 regions, while in the rest a broad line is only detected in one Balmer emission line (HFLS3$\_$g1, COS3018, MACSJ0416-Y1 and B14). In the left panel of Fig.~\ref{fig:extinction_metallicity} we show a comparison between $A_{\rm V, narrow}$ and $A_{\rm V, broad}$, with a dashed line demarcating the one-to-one relation (in this plot we exclude the candidate outflows in HZ10$\_$E and B14). We find that the dust attenuation is generally higher in the outflowing gas, where it reaches values as high as $\sim2.6$~mag, while only two galaxies display higher attenuation in the host galaxy. On average, the dust attenuation in the outflowing gas is $\sim0.59$~mag higher than in the systemic component (host galaxy). We also include in this plot the dust attenuation measured in the two ionised outflows identified by \citet{crespo-gomezRIOJADustyOutflows2025} in a star-forming galaxy at $z\sim6.1$ and in a starburst galaxy at $\sim6.2$ by \citet{marques-chavesExtremelyUVbrightStarbursts2025} where the outflowing gas also displays higher attenuation.

To constrain the gas metallicities we use standard metallicity indicators based on optical emission lines such as R2 (\OIIlines / \Hb), R3 (\OIIIb / \Hb) and N2 (\NIIb / \Ha). We separately estimate the gas metallicity in the narrow (systemic) and broad (outflow)  kinematic components when a broad component is detected in enough emission lines to compute, at least, two metallicity indicators. Such a requirement is satisfied in seven galaxies in our sample for R3 and N2. In the other systems, we compute the metallicity using the line fluxes from the single kinematic component fit. Line fluxes are corrected for dust attenuation when the indicators use emission lines with large wavelength separation (e.g. R2). Oxygen abundances are computed using the recent calibrations based on star-forming galaxies at $z=1.4-7.2$ for local galaxies from the AURORA survey \citet{sandersAURORASurveyHighRedshift2025} for the two indicators, which are combined to find a common value and provide the associated uncertainties. The right panel of Fig.~\ref{fig:extinction_metallicity} presents a comparison of the oxygen abundance in the narrow (systemic) and broad (outflow) components, with a dashed line demarcating the one-to-one relation. All seven galaxies for which this comparison can be done are at $z<6.5$. Remarkably, all the points lie on top or above the one-to-one relation, indicating that the outflowing gas is more metal-enriched than the gas in the host galaxies. On average, outflowing gas presents $\sim0.13$~dex higher oxygen abundances. We discuss the implications of these findings in Sect.~\ref{subsec:discussion_metals}.

\subsubsection{Outflow kinematics}
\label{subsec:constraining_kinematics}

In this section we explore the variation of the kinematic properties of the outflowing gas as a function of the properties of the host regions. From the spectral modelling of the integrated spectra of the outflows we derive the Full Width at Half Maximum of the broad component, $FWHM_{\rm broad}$, and the velocity difference between the broad and narrow kinematic components, $\Delta$V~=~V$_{\rm broad}$-V$_{\rm narrow}$, which are used to define the outflow velocity as V$_{\rm out}=|\Delta V| + FWHM_{\rm broad}/2$, following \citet{xuStellarAGNDrivenOutflows2025a}. To estimate the total, attenuation-corrected, \Ha luminosity of the regions hosting the outflows (L$_{\rm H\alpha, out}$) we use total line fluxes including all kinematic components, thus assuming that the outflow emission is due to photons escaping from the star-forming regions and ionising the outflowing gas, rather than in-situ star formation within the outflow. Since at this stage we are working with the spectra extracted from the outflowing regions defined in Fig.~\ref{fig:sigma_broad_maps} (red contours), the derived L$_{\rm H\alpha, out}$ values differ from those estimated in Sect.~\ref{sec:general_properties} where different apertures were used to characterise the sample. We derive the SFRs of the region hosting the outflow (SFR$_{\rm out}$) from the total L$_{\rm H\alpha}$ as in Sect.~\ref{subsec:main_sequence} and compute the \Ha luminosity surface density (thus $\Sigma_{\rm SFR, out}$) using the outflow sizes estimated in Sect.~\ref{subsubsec:size_determination}. It is worth noting here the advantage of using the spatially resolved information provided by the NIRSpec IFS observations to perform an adequate estimation of $\Sigma_{\rm SFR, out}$. We also explore variations as a function of the stellar mass of the host galaxies/regions using the values obtained in Sect.~\ref{sec:analysis} for the large apertures (pink) defined in Fig.~\ref{fig:integrated_oiii_maps}.    

\begin{figure*}
\centering
\includegraphics[width=0.98\textwidth]
{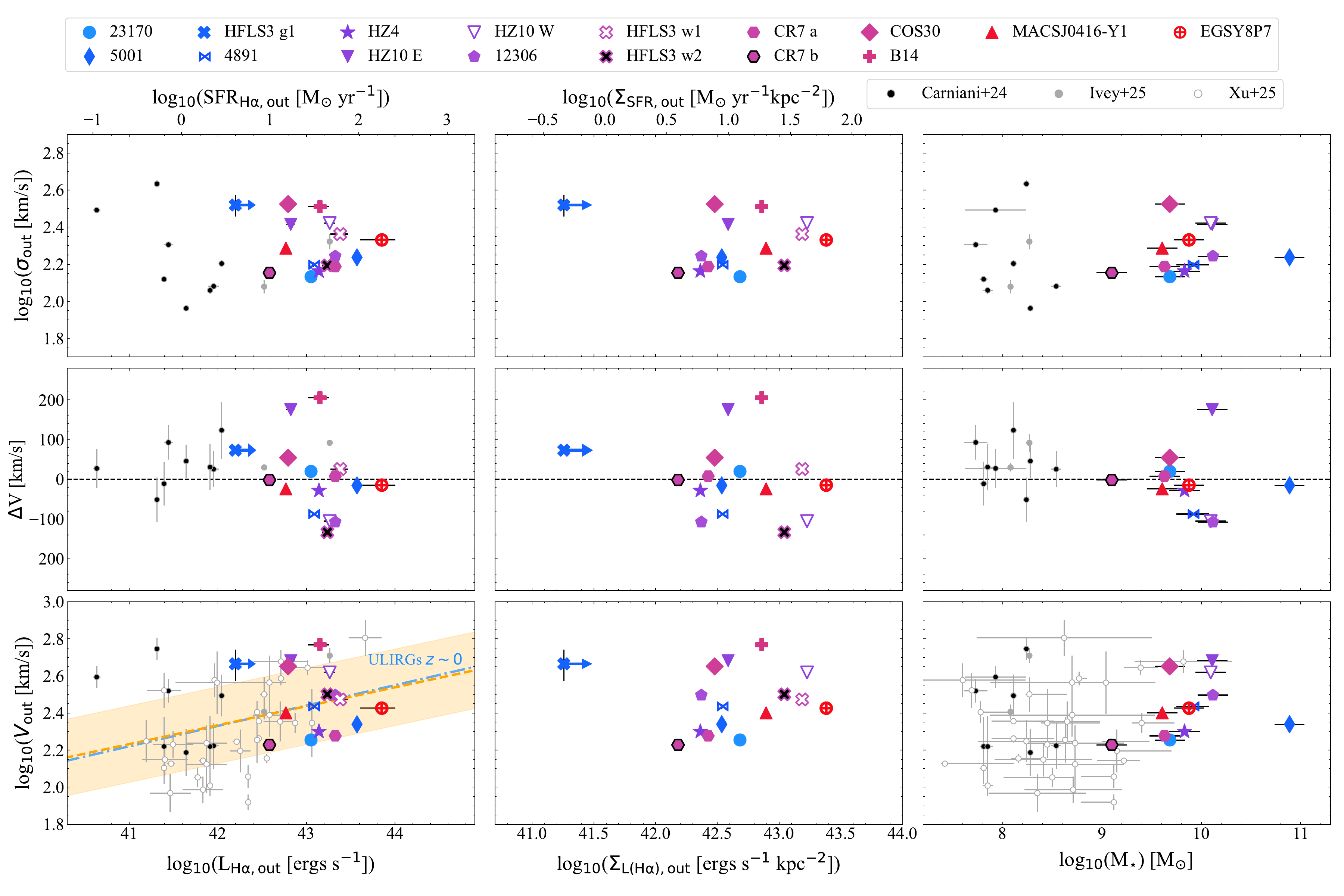}
\caption{Kinematic properties of the outflows, $\sigma_{\rm out}$, $\Delta$$V$, V$_{\rm out}$, as a function of the properties of the regions hosting them: L$_{\rm H\alpha, out}$ (SFR$_{\rm out}$), $\Sigma$$_{\rm L(H\alpha), out}$ ($\Sigma_{\rm SFR, out}$), and M$_{*}$. Large symbols correspond to the outflows identified in our study and are colour-coded according to their redshifts, as in Fig.~\ref{fig:extinction_metallicity}. Results from previous studies of ionised outflows at $3<z<9$ are shown as small black, grey, and white circles \citep{carnianiJADESIncidenceRate2024, iveyExploringSpatiallyResolved2026, xuStellarAGNDrivenOutflows2025a}. In the bottom-left panel we show as an orange dashed line, the best fit to the data points together with a shaded region marking the 1$\sigma$ scatter. The data presents a statistically significant ($>3\sigma$) positive correlation. For comparison, in that panel we also show as a dot-dashed blue line the correlation found for local ULIRGs by \citet{arribasionizedGasOutflows2014}.} 
\label{fig:kinematic_properties}
\end{figure*}

In Fig.~\ref{fig:kinematic_properties} we show a comparison between the kinematic properties of the outflow ($\sigma_{\rm out}$, $\Delta$V, V$_{\rm out}$) and the properties of the host. For comparison, we also include in the plots the samples of outflows from \citet{carnianiJADESIncidenceRate2024}, \citet{iveyExploringSpatiallyResolved2026} and \citet{xuStellarAGNDrivenOutflows2025a}. At a first glance, these plots highlight that our sample of galaxies covers higher values of SFR and stellar mass than previous studies. Thus, this expanded parameter space allowed us to conduct a comprehensive census of galactic outflows at $3<z<9$ across a wide a range of stellar masses and SFRs. 

As shown in the plots in the top row of Fig.~\ref{fig:kinematic_properties}, our sample of ionised outflows display velocity dispersions, traced by $\sigma_{\rm out}$, within $\sim130-340$~km/s, similar to those measured in less luminous and less massive star-forming galaxies in previous works. The velocity dispersion does not seem to vary as a function of the host properties within the wide ranges we probe. The distribution of the $\Delta V$ values presented in the middle panels illustrates that most of the outflows we identify are blue-shifted with respect to the systemic emission or present slightly red-shifted, positive velocity differences $<100$~km/s. There are only two outflows in our sample with large, positive velocities $\sim200$~km/s, HZ10$\_$E and B14. As explained in Sect.~\ref{sec:outflow_regions}, we consider these outflows as `candidates', since in a standard outflow scenario we would expect to see the unattenuated blue-shifted gas while the redshifted emission is more likely to be obscured by the host galaxy. Nevertheless, as shown in the plots, we note that red-shifted outflows are also reported in the previous works by \citet{carnianiJADESIncidenceRate2024} and \citet{iveyExploringSpatiallyResolved2026}, suggesting that the geometry and dust distribution can frequently make the red-shifted outflowing gas more prominent than the blue-shifted counterpart. Red-shifted broad components can also be explained by gas inflows of pristine gas. However, our results of a higher metallicity in the outflowing gas (see Sect.~\ref{subsec:results_av_z}) do not support this scenario. 

Finally, the outflows identified in this study present outflow velocities V$_{\rm out}\sim170-600$~km~s$^{-1}$. It is noticeable that at high SFRs ($>1.5$ M$_{\odot}$yr$^{-1}$) and high stellar masses (log$_{10}$~(M$_\star$/M$_\odot$)$>9.5$), regimes mostly covered by our sample and by that from \citet{xuStellarAGNDrivenOutflows2025a}, the outflows present V$_{\rm out}>150-200$~km/s, while outflows at lower SFRs and stellar masses span towards lower velocities, down to 100~km~$^{-1}$. Moreover, supported by these different velocity regimes, we identify a trend of increasing V$_{\rm out}$ as a function of L$_{\rm H\alpha, out}$ including all samples, with a Spearman’s correlation test probability $>3\sigma$. The best linear fit to the data points (with a slope of $\sim0.098$) is shown (orange, dashed line) in the bottom-left panel of Fig.~\ref{fig:kinematic_properties}. We also show the relation found for local ULIRGs (blue, dot-dashed line) by \citet{arribasionizedGasOutflows2014} shifted to match our relation at L$_{\rm H\alpha, out}=10^{42.5}$~L$_{\odot}$, which is obtained within a similar range in SFRs using also \Ha, although with a slightly different definition of V$_{\rm out}$\footnote{The slope of the relation reported in \citet{arribasionizedGasOutflows2014} remains the same using both definitions of V$_{\rm out}$.}. Remarkably, the two correlations have a very similar slope indicating that, regardless of the cosmic time, there is an expected range in the velocity of ionised outflows at a given SFR. We do not find statistically significant correlations between $\sigma_{\rm out}$, V$_{\rm out}$ and the host properties $\Sigma$$_{\rm L(H\alpha), out}$ nor M$_{*}$. Although not shown in the plot, we also explore correlations between the kinematics and the specific SFR, finding no statistically significant trends. 

\section{Discussion}
\label{sec:Discussion}

\subsection{High prevalence of ionised outflows}
\label{subsec:incidence_outflows} 
We have identified ionised outflows in 14 out of the 40 individual star-forming galaxies/regions explored in our sample of 15 complex systems. Compared to previous works covering similar redshifts (3<z<9), this incidence of outflows ($\sim35\%$) is similar to that reported by \citet[][25-40\% in the sample of 52 galaxies]{carnianiJADESIncidenceRate2024}, but higher than in the works by \citet[][$3.4\%$ (34/1007)]{cooperHighvelocityOutflowsOIII2025} and \citet[][23\% (30/130)]{xuStellarAGNDrivenOutflows2025a}. These differences can be ascribed to several factors, including (1) the stellar mass coverage (log$_{10}$~(M$_\star$/M$_\odot$)~=~$8.5-11$), (2) the type and depth of the observations and (3) the outflow selection criteria. (1) Regarding stellar mass, our sample of galaxies with identified outflows have stellar masses within log$_{10}$~(M$_\star$/M$_\odot$)$=8.75-10.9$, while previous works generally contain galaxies with log$_{10}$~(M$_\star$/M$_\odot$)$<9.5$. Given that the incidence of outflow has been observed to increase with stellar mass at low \citep[e.g.][]{rodriguezdelpinoPropertiesionizedOutflows2019} and high redshift \citep[e.g.][]{carnianiJADESIncidenceRate2024}, it is expected that our sample presents a higher incidence than previous works that studied lower mass systems. (2) The spatial resolution of our IFS observations allowed us to isolate the regions potentially hosting ionised outflows, extract all the emission from the outflows that might be missed in MOS observations, and minimise contamination from mergers that might present similar spectral signatures. However, with our data is less challenging and subject to lower uncertainties than for the NIRCam Wide Field Slitless and NIRSpec MSA observations used in the works mentioned above. Moreover, the high spectral resolution (R2700) of our observations (with the exception of B14 that was observed with R1000) also enhances the detectability of a possible broad component. This result is demonstrated by the increasing fraction of outflows (from 23\% to 30\%) identified in \citet{xuStellarAGNDrivenOutflows2025a} when only high resolution observations are considered and by the low incidence of outflows reported in \citet{cooperHighvelocityOutflowsOIII2025} where only R1000 data was employed. (3) The outflow selection criteria in individual studies, including ours (see Sect.~\ref{sec:outflow_regions}), generally require a minimum S/N (e.g. $>3$) in the broad component and a statistical improvement in the spectral modelling that favours a secondary kinematic component associated with an outflow. Although each study applies different thresholds to these criteria, they are expected to be reasonably consistent. However, a direct comparison of the criteria would be needed to properly quantify any potential differences, but this is beyond the scope of this work. 

\begin{figure*}
    \centering
    \includegraphics[width=0.8\textwidth]{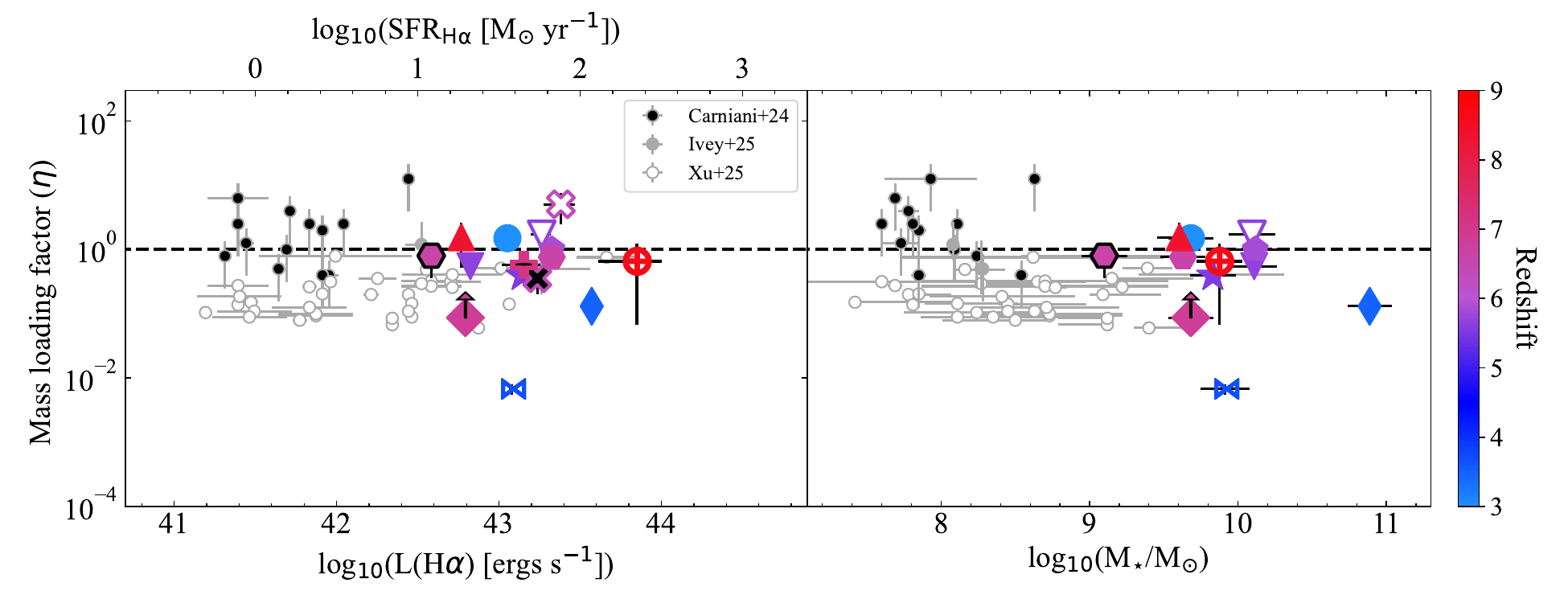}
    \caption{Mass loading factor ($\eta$) of the outflows in our sample as a function of SFR$_{\rm H\alpha}$ (L$_{\rm H\alpha}$; left) and stellar mass, M$_\star$, of the host galaxy (right). Symbols are the same as in Fig.~\ref{fig:kinematic_properties}. For comparison, we show the sample of star-forming galaxies with outflows at $3<z<9$ from \citet{carnianiJADESIncidenceRate2024}, \citet{xuStellarAGNDrivenOutflows2025a} and \citet{iveyExploringSpatiallyResolved2026}. }
    \label{fig:mass_loading}
\end{figure*}

The detection of ionised outflows in most of our systems resembles the results from studies of local ULIRGs where ionised outflows are ubiquitous \citep[e.g.][]{bellocchiVLTVIMOSIntegral2013, arribasionizedGasOutflows2014} due to their high masses (log$_{10}$~(M$_\star$/M$_\odot$)$>9.5$) and SFRs ($>1.5$ M$_{\odot}$yr$^{-1}$). In addition, similar to our sample of star-forming galaxies, local ULIRGs also tend to display high disordered motions and signatures of ongoing mergers \citep{crespogomezStellarKinematicsNuclear2021, pernaPhysicsULIRGsMUSE2022a}. Finally, our results are also comparable to the widespread presence of outflows found in AGNs at $z\sim3$ \citep{bertolaGANIFSMapping352025, venturiGANIFSPowerfulFrequent2025} and at $z>4$ \citep{marshallGANIFSBlackHole2023, liuFastOutflowHost2024, loiaconoQuasargalaxyMerger622024, suhSuperEddingtonaccretingBlackHole2025} also with NIRSpec IFS observations.

\subsection{Impact on the host galaxy}
\label{subsec:discussion_impact} 

Although starburst-driven outflows are expected to have a lower impact on the host galaxies than AGN-driven ones, they might still affect their star formation activity \citep[e.g.][]{carnianiJADESIncidenceRate2024} and play an important role in the regulation of the metal content in galaxies \citep[e.g.][]{chisholmMetalenrichedGalacticOutflows2018a}. In this section we explore these two scenarios. 

\subsubsection{Limited ejective feedback in high-z star-forming systems}
\label{subsec:discussion_sf_suppression} 

A common approach to assessing whether outflows are suppressing star formation in their host galaxies is to compare the outflowing mass rate, $\dot{\rm M}_{\rm out}$, that is the rate at which gas is expelled from the star-forming region, with the SFR$_{\rm out}$, which quantifies the rate at which gas is converted into stars. The ratio between these two quantities is known as the mass loading factor, $\eta$~=~$\dot{\rm M}$$_{\rm out}$/SFR$_{\rm out}$. A value of $\eta>1$ is interpreted as an indication that the outflow is suppressing star formation in the host. To estimate the outflow rate, we assumed an outflow with a constant mass outflow rate with time \citep{lutzMolecularOutflowsLocal2020}, which is given by

\begin{equation}
\dot{M}_{\rm out}= M_{\rm out} V_{\rm out} / r_{\rm out},
\label{eq:rate}
\end{equation}

\begin{figure}
\centering
\includegraphics[width=0.45\textwidth]
{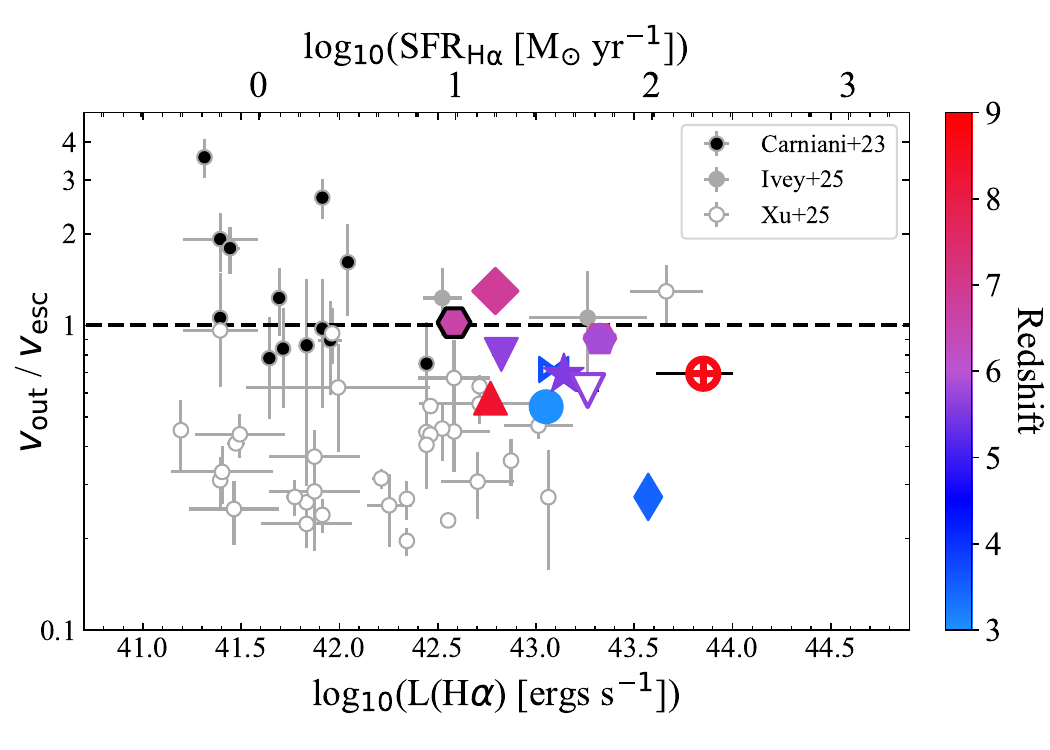}
\caption{Ratio between the outflow velocity (V$_{\rm out}$) and the V$_{\rm esc}$ as a function of SFR$_{\rm H\alpha}$ (L$_{\rm H\alpha}$).}
\label{fig:vesc_mass}
\end{figure}

where M$_{\rm out}$ is the mass of the outflowing gas and r$_{\rm out}$  is the size of the outflow \citep[e.g.][]{maiolinoEvidenceStrongQuasar2012, gonzalez-alfonsoMolecularOutflowsLocal2017}. We continued by estimating M$_{\rm out}$ following the method outlined in \citet{carnianiJADESIncidenceRate2024}, specifically their Eq. 4, which we reproduce below: 

\begin{equation}
M_{\rm out} = 0.8\times10^8 \left( \frac{L^{\rm corr}_{\rm [OIII], out}}{\rm 10^{44}~erg~s^{-1}}\right)\left( \frac{Z_{\rm out}}{\rm Z_{\odot}}\right)^{-1}\left( \frac{n_{\rm out}}{\rm 500~cm^{-3}}\right)^{-1}~{\rm M_{\odot}}.
\label{eq:moutoiii}
\end{equation} 

\noindent The $L^{\rm corr}_{\rm [OIII], out}$ is the \OIII luminosity of the broad component corrected for dust attenuation whereas $Z_{\rm out}$ and $n_{\rm out}$ are the metallicity and electron density of the outflow, respectively.  We perform the correction for dust attenuation in the broad component using the $A_{\rm V, broad}$ values estimated in Sect.~\ref{subsec:results_av_z}. In the four cases (HFLS3$\_$g1, COS30-18, B14 and MACSJ0416-Y1) where $A_{\rm V, broad}$ could not be estimated, we assume that the dust attenuation in the outflow is the same as in the host. Regarding the metallicity of the outflow, we use individual estimates for the galaxies where we could measure it directly from the spectra (see Sect.~\ref{subsec:results_av_z}), while for the rest of systems we assume for the outflow the same gas metallicity estimated for the systemic component. Finally, an accurate determination of the electron density of the outflow remains challenging, primarily because the \OII and \SII emission line doublets required for such measurements are frequently blended or display insufficient S/N. Due to this difficulty, we consider the same electron density for the outflow and systemic components. Given that previous works have estimated the electron density of the host galaxies we employ those values when available: 540~cm$^{-3}$ for 5001 \citep{lampertiGANIFSJWSTNIRSpec2024a}; 776~cm$^{-3}$ for 4891 \citep{rodriguezdelpinoGANIFSCoevolutionHighly2024}; 270~cm$^{-3}$ for HZ4 \citep{parlantiGANIFSMultiphaseAnalysis2025}; 1000~cm$^{-3}$ for HZ10\_E and 316~cm$^{-3}$ for HZ10\_W \citep{jonesGANIFSWitnessingComplex2025}; 1200~cm$^{-3}$ for COS3018 \citep{scholtzGANIFSISMProperties2025}; 2200~cm$^{-3}$ for EGSY8P7 \citep{zamoraGANIFSUnderstandingionization2025}. For the remaining nine host galaxies/regions we adopt a common value of 500~cm$^{-3}$ based on the measured electron densities within $3<z<9$ in \citet[][ see also \citealt{marconciniGANIFSInterplayMerger2024a}]{isobeRedshiftEvolutionElectron2023a}. Under these assumptions, we used Eq. 2 to estimate the mass of the outflowing gas for the whole sample, which is found to range between $\sim10^6$ and $\sim9\times10^8$ M$_{\odot}$.

With these estimations of the outflowing gas mass together with the values derived for the outflow velocity, V$_{\rm out}$, in Sect.~\ref{subsec:constraining_kinematics} and for the outflow radius, r$_{\rm out}$, in Sect.~\ref{subsubsec:size_determination}, we calculate the  mass outflow rate, $\dot{\rm M}_{\rm out}$, using Eq. 1, and the mass loading factor $\eta$. The estimated values of $\eta$ for the outflows identified in this work are listed in Table~\ref{table:outflow_properties} and shown in Fig.~\ref{fig:mass_loading} as a function of the SFR$_{\rm H\alpha}$ (derived from the \Ha luminosity, L$_{\rm H\alpha}$; left) and the stellar mass of the hosts (right) in comparison with other works at similar redshifts \citep{carnianiJADESIncidenceRate2024, xuStellarAGNDrivenOutflows2025a}. In our sample, $\eta$ ranges from $<10^{-3}$ to $5$, with five out the 14 outflows presenting values equal to or above unity. These results suggest that in these five systems, outflows may be actively suppressing star formation in the host galaxies, whereas in the rest of the sample, their impact on star formation appears to be less significant. The generally low values of $\eta$ found in our sample of outflows are in good agreement with the findings reported in \citet{xuStellarAGNDrivenOutflows2025a}, where they adopt the same definition of V$_{\rm out}$ as in our work and apply the redshift-dependent scaling relation from \citet{isobeRedshiftEvolutionElectron2023a}, assuming n$_{\rm e}=1000$~cm$^{-3}$ at $z=8$. On the contrary, the values of $\eta$ measured in \citet{carnianiJADESIncidenceRate2024} are significantly higher, with a high fraction of their outflows (ten out of 14) displaying $\eta>1$. A significant fraction of outflows with $\eta>1$, especially at low stellar masses (log$_{\rm 10}$~(M$_\star$/M$_\odot$)$<9$), has also been reported by \citet{cooperHighvelocityOutflowsOIII2025}. We ascribe these discrepancies to the lower electron densities (n$_{\rm e}=380$~cm$^{-3}$) adopted in these works and to the fact that their definition of V$_{\rm out}$ ($\Delta$V + $2\sigma_{\rm out}$ instead of $\Delta$V + FWHM$_{\rm out}/2$ as in our work) leads to higher velocities of the outflowing gas and, consequently, to higher values of $\eta$. If we assumed the same lower electron density for the outflows in our sample that lack an individual measurement of the host $n_{\rm e}$, $\eta$ would increase by $30\%$, having two more outflows (CR7\_a, b) with $\eta>1$. Alternatively, if we used the redshift-dependent scaling relation from \citet{isobeRedshiftEvolutionElectron2023a} as in \citet{xuStellarAGNDrivenOutflows2025a}, which implies higher electron densities than our adopted value (n$_{\rm e}=500$~cm$^{-3}$), the outflows with $\eta>1$ in our work remain unchanged. These differences indicate that adopting electron densities lower than those typical of high-redshift galaxies can lead to an overestimation of the mass-loading factor. Finally, it is important to bear in mind that, although our results suggest that the starburst-driven outflows do not have a strong impact in the star formation activity of the hosts, these estimates only account for the impact of the ionised, and the actual impact might be larger if the neutral and molecular phases are considered \citep[e.g.][]{jonesDetectionHighredshiftMolecular2019, daviesJWSTRevealsWidespread2024}.

\subsubsection{Dust and metal redistribution through galactic outflows}
\label{subsec:discussion_metals} 
The analysis presented in Sect.~\ref{subsec:results_av_z} and shown in Fig.~\ref{fig:extinction_metallicity} confirms a result that was already glimpsed in works on individual galaxies from our sample \citep{lampertiGANIFSJWSTNIRSpec2024a, rodriguezdelpinoGANIFSCoevolutionHighly2024, parlantiGANIFSMultiphaseAnalysis2025}, but now has been established analysing a large sample: In high-redshift ($3<z<9$) star-forming galaxies, outflowing gas has generally higher dust attenuation and oxygen abundances than the gas in the host galaxies. The higher dust attenuation is observed in eight out of 11 outflows, while a higher or roughly  equal metallicity to the host is found in all seven outflows that could be evaluated. Interestingly, recent works by \citet{crespo-gomezRIOJADustyOutflows2025} and \citet{ marques-chavesExtremelyUVbrightStarbursts2025} have also found higher dust attenuation in the outflowing gas in two galaxies at $z\sim6.1-6.2$, supporting the general behaviour we observe in our sample. Metal-enriched outflows have also been found in large samples of local star-forming galaxies \citep[e.g.][]{rodriguezdelpinoPropertiesionizedOutflows2019}, and their influence has been suggested to contribute to the shaping of the mass-metallicity relation based on the observed decrease in the metal-loading factor as a function of stellar mass \citep{chisholmMetalenrichedGalacticOutflows2018a}. Nevertheless, although metal-enriched outflows are common in the local Universe, to our knowledge, this is the first time this result is reported in a relatively large sample of galaxies at $3<z<9$.

Finally, the higher metallicity of the gas associated with the broad component emission also confirms the outflowing nature of this component, thus discarding an inflow or merger scenario leading to the broad emission. In the former case, the metallicity of the gas coming from the circumgalactic medium (CGM) is expected to be lower, as it has not been processed by star formation \citep[e.g. ][]{arribasGANIFSCoreExtremely2024a}. In the latter, the metallicity is not expected to be systematically high in one of the two components. We also note that, even though the high metallicities in the outflowing gas could be due to the enhancement in N2 expected in shocks \citep[e.g.][]{hoSAMIGalaxySurvey2014}, we do not expect this effect to be relevant since we require an additional metallicity indicator, which is generally R3, to reduce degeneracies in the estimations of the oxygen abundances.

We evaluate now whether the dust and metal-enriched outflows in our galaxies could be expelled from their host galaxies and reach the CGM by comparing the maximum velocity of the outflowing gas, V$_{\rm out}$, with the escape velocity, V$_{\rm esc}$, of the galaxy. The estimation of V$_{\rm esc}$ can be obtained from the dynamical masses of the galaxies following the method described in \citet{arribasionizedGasOutflows2014}. In our case, since we do not have estimations of the dynamical masses, we assume M$_{\rm dyn}=2$~$\times$~M$_{\star}$, based on the values obtained for individual galaxies in our sample \citep[e.g.][]{rodriguezdelpinoGANIFSCoevolutionHighly2024}. Then, we estimate the escape velocity at the radius of the outflow, r$_{\rm out}$, for an isothermal sphere truncated at 10 kpc. To estimate whether part of the gas could escape the potential well of their host galaxies, we plot in Fig.~\ref{fig:vesc_mass} the ratio between V$_{\rm out}$ and V$_{\rm esc}$ and explore its variation with SFR. In only one case, COS30, the outflowing gas presents enough velocity to escape the host galaxies and reach the CGM. In the rest of the cases, the outflowing gas remains bound, which is consistent with the low mass loading factor inferred above and further corroborate the idea that such outflows are not powerful enough to alter the host equilibrium by suppressing SF. Instead, they probably redistribute gas throughout the galaxies, potentially contributing to the observed gradients and heterogeneous distribution of dust and metals in high redshift galaxies \citep[e.g.][]{venturiGasphaseMetallicityGradients2024}. In the interpretation of our results, we note that in compact/rotating systems where M$_{\rm dyn}$ can exceed $2\times$M$_{\star}$ we might underestimate V$_{\rm esc}$. However, this would only strengthen our conclusion that the outflows remain bound. These results are in contrast with the detection of metal-enriched gas from \CII emission extended on physical sizes of $\sim20-30$~kpc (diameter scale) firstly identified by \citet{fujimotoFirstIdentification102019} in stacking analysis at $5<z<7$ and later confirmed in a larger sample by \citet{ginolfiALPINEALMAIISurvey2020} at $4<z<6$.   

\section{Summary and conclusions}
\label{sec:conclusions}

In this work, we have presented a search for and characterisation of ionised outflows in a sample of 15 complex star-forming systems comprising 40 individual galaxies/regions with no evidence of AGN activity using JWST/NIRSpec IFU data from the GA-NIFS GTO program. We used the \OIII and \Ha emission lines as tracers of ionised gas to identify broad kinematic components indicative of potential outflows. Crucially, the spatially resolved information provided by the NIRSpec IFU allowed us to isolate the regions hosting these outflows and enabled more reliable estimates of their physical properties and impact on the host galaxies than those obtained from integrated spectra alone. Our study complements previous work on ionised outflows at $3<z<9$ by probing more luminous and massive systems, which allowed us to perform a comprehensive census of galactic outflows within a wide range of SFRs (log$_{\rm 10}$(SFR$_{\rm H\alpha}$/M$_\odot$yr$^{-1}$)~=~$[-0.11,2.5]$) and stellar masses (log$_{10}$~(M$_\star$/M$_\odot)$~$=7.5-11$). We used the integrated spectra from the regions hosting the outflows to derive the kinematics of the outflowing gas as well as its dust attenuation and oxygen abundance. In addition, we evaluated whether the detected outflows might contribute to the suppression of star formation in the host galaxies and/or expel gas out to the CGM. The main findings of these analyses are explained in the following: 

\begin{itemize}  
\item We identified outflows in 13 out of 15 star-forming systems. In terms of the individual galaxies/regions explored, we found outflows in 14 out of 40 of them, which translates into an outflow incidence of 35\%. We ascribe this high prevalence of ionised outflows, which is larger than in previous works at similar redshifts, mainly to the higher stellar masses (log$_{\rm 10}$~(M$_\star$/M$_\odot$)$>8.5$) and SFRs (log$_{\rm 10}$(SFR$_{\rm H\alpha}$/M$_\odot$yr$^{-1}$)~=~$[-0.3,2.5]$) of our galaxies and the advantages provided by our deep IFS data to identify galactic outflows. 

\item The outflowing gas presents, on average, $\sim0.59$~mag higher dust attenuation and $\sim0.13$~dex higher oxygen abundances than the host galaxies, indicating that outflows entrain significant amounts of dust and metals. However, the velocities are not large enough (V$_{\rm out}$/V$_{\rm esc} < 1$) for the outflowing gas to escape the host galaxies and reach the CGM, and instead they probably redistribute the gas throughout the galaxies.

\item The outflows identified in this study present velocity dispersions within $\sigma_{\rm out}=130-340$~km/s, similar to those measured in less luminous and massive star-forming galaxies in previous works. We identified a statistically significant trend of increasing V$_{\rm out}$ as a function of SFR across the wide range covered by this and previous samples of star-forming galaxies at $3<z<9$.

\item The majority of the ionised outflows we detected (9/14) do not seem to have a relevant impact on the star formation activity based on the values obtained for the mass-loading factor ($\eta$~$<$~$1$). Comparison with previous works indicates that, considering similar assumptions in electron densities and the same definition of the outflow velocity as in our work, starburst-driven outflows have a limited impact on the star formation activity of the hosts within a wide range of SFRs (log$_{\rm 10}$(SFR$_{\rm H\alpha}$/M$_\odot$yr$^{-1}$)~=~$[-0.11,2.5]$) and stellar masses (log$_{10}$~(M$_\star$/M$_\odot)$~$=7.5-11$). 

\end{itemize}

This work represents the first spatially resolved study of star formation--driven galactic outflows in a relatively large sample of galaxies at $3<z<9$ and provides a detailed characterisation of their properties and impact on the host galaxies. Our results suggest that massive galaxies at high redshift are potential hosts of ionised outflows that might not have a significant impact on the suppression of star formation or the expelling of gas to the CGM but could potentially redistribute metals and dust throughout galaxies. Given that this work has focused only on the ionised phase of the gas, a more complete picture would be provided by the study of the neutral and molecular phases. 


\begin{acknowledgements}
We acknowledge the referee for their useful comments and suggestions that have contributed to improve this manuscript. We would like to acknowledge the whole JWST mission and the instrument science teams for a successful launch and commissioning of the observatory. 
BRP, SA, MP acknowledge support from the research projects PID2021-127718NB-I00, PID2024-159902NA-I00, PID2024-158856NA-I00, and RYC2023-044853-I of the Spanish Ministry of Science and Innovation/State Agency of Research (MCIN/AEI/10.13039/501100011033), FSE+, and by “ERDF A way of making Europe''.
IL acknowledges support from PRIN-MUR project “PROMETEUS”  financed by the European Union -  Next Generation EU, Mission 4 Component 1 CUP B53D23004750006.
AJB acknowledges funding from the ``First Galaxies'' Advanced Grant from the European Research Council (ERC) under the European Union’s Horizon 2020 research and innovation programme (Grant agreement No. 789056).
SC, GV and SZ acknowledge support by European Union’s HE ERC Starting Grant No. 101040227 - WINGS.
MP, GC and EB acknowledge the support of the INAF Large Grant 2022 "The metal circle: a new sharp view of the baryon cycle up to Cosmic Dawn with the latest generation IFU facilities". 
GC and EB also acknowledge the INAF GO grant ``A JWST/MIRI MIRACLE: Mid-IR Activity of Circumnuclear Line Emission''. EB acknowledges funding through the INAF ``Ricerca Fondamentale 2024'' programme (mini-grant 1.05.24.07.01).
H\"U acknowledges funding by the European Union (ERC APEX, 101164796). Views and opinions expressed are however those of the authors only and do not necessarily reflect those of the European Union or the European Research Council Executive Agency. Neither the European Union nor the granting authority can be held responsible for them.
FDE, RM, GCJ, and JS acknowledge support by the Science and Technology Facilities Council (STFC), by the ERC through Advanced Grant 695671 ``QUENCH'', and by the UKRI Frontier Research grant RISEandFALL.
This work is based on observations made with the NASA/ESA/CSA James Webb Space Telescope. The data were obtained from the Mikulski Archive for Space Telescopes at the Space Telescope Science Institute, which is operated by the Association of Universities for Research in Astronomy, Inc., under NASA contract NAS 5-03127 for JWST. 

\end{acknowledgements}

\bibliographystyle{aa}
\bibliography{GA-NIFS_SF_outflows}

\begin{appendix}

\section{Aperture corrections}
\label{app:aperture_corrections}
\begin{figure}[h]
\centering
\includegraphics[width=0.4\textwidth]
{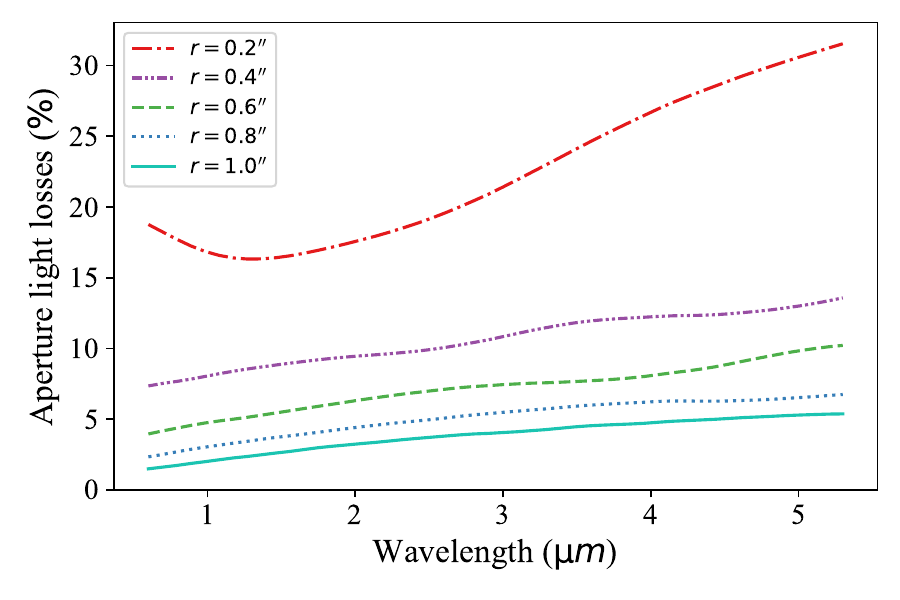}
\caption{Light losses associated with the extraction of fluxes from a simulated data cube of the NIRSpec PSF in IFS mode, using circular apertures with different radii (see Appendix~\ref{app:aperture_corrections}).}
\label{fig:aperture_losses}
\end{figure}

In the extraction of integrated fluxes from the data cubes (Sects. \ref{sec:general_properties} and \ref{sec:outflow_regions}) we apply a correction for the light losses outside the adopted apertures caused by the NIRSpec PSF. To compute these losses, we start by generating a data cube containing the simulated PSF of the NIRSpec IFS mode using the \textsc{STPSF}\footnote{https://stpsf.readthedocs.io/} Python package. At this stage we work with a data cube generated in the R100 mode, which allows the losses across the entire wavelength range of NIRSpec to be computed and provides exactly the same results as the data cubes of individual filter and grating configurations in medium and high resolution modes. We performed a spectral binning of the PSF data cube in windows of 50 pixels and computed the fraction of the total light contained within each aperture. Corrections are estimated as the inverse of this fraction and are obtained at each wavelength through interpolation. As an example, in Fig.~\ref{fig:aperture_losses} we show the light losses as a function of wavelength corresponding to circular apertures with varying radius from 0.2$^{\arcsec}$ to 0.1$^{\arcsec}$. For small apertures, $r\leq0.2\arcsec$ the light losses become significant, reaching $20-30\%$ of the total light, while at larger apertures ($r\geq0.6\arcsec$) losses are always lower than $10\%$. These results are comparable to those reported by \citet[][Fig. B.1]{zamoraGANIFSHighlyOverdense2025} using a QSO to model the PSF of NIRSpec.

\vspace{-0.65cm}

\section{Identification of outflows in individual systems}
\label{app:individual_systems}

\subsection{GS-23170}

This object is included in the VANDELS survey (\citealt{mclureVANDELSESOPublic2018}, \citealt{pentericciVANDELSESOPublic2018}) and studied by \cite{saxenaPropertiesHeIIl16402020}, who identified it as a bright and "narrow" (FWHM~$\sim450$~km/s) \HeIIline emitter . The presence of an AGN was excluded on the basis of the missing \CIV and X-ray emission. 

The present NIRSpec-IFS data reveal that \OIIIb emission from this galaxy is characterised by the presence of two bright cores (23170-N and 23170-S) within a projected distance of $\sim2-3 $~kpc. The emission in the northern core is well modelled with two kinematic components, with $\sigma_{\rm broad}>150$~km/s within a radius of $\sim1.5$~kpc around the peak in the continuum emission, consistent with an outflow scenario. The southern core also presents two kinematic components, although in this case their velocity dispersions are low ($\sigma < 100$~km/s), suggesting the overlap along the line of sight of two kinematically distinct structures (e.g. a clump and the extended disc of component N). 
More to the south there is another region with high velocity dispersion, although in this case only one kinematic component is needed to model the data. Since this region is not directly connected to the outflow at the north,  and its continuum emission is weak, we associate its turbulence with a tidal tail or interaction rather than to an outflow. 

\subsection{GS-5001}

The GS-5001 system lies within the central regions of a large-scale overdensity of galaxies that has been classified as a candidate protocluster \citep{franckCandidateClusterProtocluster2016}.
The central and brightest source (hereafter 5001-C) and its close environment have been recently studied by means of NIRSpec-IFS data by \cite{lampertiGANIFSJWSTNIRSpec2024a}. These observations cover a region of about 20 kpc, which, in addition of 5001-C,  also includes two other close companions at similar redshifts, GS-4923 (we refer to it as 5001-S) and 5001-N,  as well as several substructures revealing clear signs of interactions\footnote{The FoV also includes GS-4923, but this galaxy has a photometric redshift of about 0.2 according to Rainbow Data Base (\url{https://arcoirix.cab.inta-csic.es/Rainbow_Database/}), and therefore it is ignored in the present study.}. In their study, \citet{lampertiGANIFSJWSTNIRSpec2024a} already identified a spatially resolved ionised bi-conical outflow with an extension of $\sim3$~kpc along the minor axis of the central galaxy of this system (see their Fig.~6). Similarly to their results, we find a region with enhanced velocity dispersion extending to the East and West from the central regions of 5001-C, reaching values of up to $300$~km/s, thus consistent with an outflow. We do not find clear evidence of outflows in 5001-S nor in 5001-N.

\subsection{GS- 4891}
By means of NIRSpec-IFS data, this system was studied in detail by \citet{rodriguezdelpinoGANIFSCoevolutionHighly2024}, who found a spatially resolved outflow extending up to a distance of $\sim1.5$~kpc from the central parts of the main galaxy (hereafter 4891-C). In our analysis, we also identify this extended emission along the Southeast -- Northwest direction, slightly smaller than in the previous work (1.05~kpc vs 1.2~kpc) due to the different criteria applied to identify outflows. Nevertheless, the overall properties of the outflow are very similar in both works. \citet{rodriguezdelpinoGANIFSCoevolutionHighly2024} also find in the FoV two companions galaxies at a redshift similar to 4891-C, 4891-N and GS-28356 (we refer to it as 4891-E), but none of these galaxies has traces of outflowing gas.

\subsection{HZ4}

HZ4 is a complex star-forming system, with an extensive set of multi-wavelength pre-JWST datasets including from \textit{Hubble} Space Telescope (HST: \citealt{scovilleCOSMOSHubbleSpace2007}), Keck \citep{malleryLyaEmissionHighredshift2012}, and ALMA (\citealt{capakGalaxiesRedshifts52015}, \citealt{herrera-camusKiloparsecViewTypical2021a}, \citealt{jonesALPINEALMAIISurvey2021}). The recently analysed NIRSpec/IFS data \citep{parlantiGANIFSMultiphaseAnalysis2025} have revealed three close merging galaxies/structures in the system, HZ4-N, HZ4-C, and HZ4-S, all within a projected distance of about 4~kpc. 

They detect and characterise ionised outflows in HZ4-C  and found that it extends over regions of up to 4~kpc with velocity dispersions of $\sim200-250$~km/s (see their Fig.~6), properties that are reproduced in our analysis. HZ4 is also the most distant star-forming
galaxy presenting evidence of a neutral outflow, traced by the \CII line measured with ALMA \citep{herrera-camusKiloparsecViewTypical2021a}. \citet{parlantiGANIFSMultiphaseAnalysis2025} compared the neutral and ionised outflows, finding that they are partially co-spatial, with the former dominating the total mass outflow rate by more than one order of magnitude.

\subsection{HZ10}

HZ10 was originally detected as a Lyman break galaxy in the 2 square degree Cosmic Evolution Survey \citep{scovilleCosmicEvolutionSurvey2007} field. Follow-up observations revealed that this source is a UV-luminous Ly$\alpha$ emitter at $z=5.659$
with bright far-infrared emission (\CII) consistent with SFR$_{\rm IR}=169$~M$_{\odot}$yr$^{-1}$ \citep{capakGalaxiesRedshifts52015}. Using NIRSpec-IFS data, \citet{jonesGANIFSWitnessingComplex2025} have recently studied this complex star-forming system and identified three emission-line galaxies within a projected distance of $\sim10$~kpc, HZ10$\_$E, HZ10$\_$C, and HZ10$\_$W, likely in a merging pre-coalescence phase. They also identify blue- and red-shifted asymmetries of the order of $\sim200$~km/s in the \OIII and \Ha emission line profiles of the East and West members, that they attributed to the possible presence of an outflow or tidal tails.

In our $\sigma_{\rm broad}$ map (Fig.~\ref{fig:sigma_broad_maps}), we find that these two regions present the highest velocity dispersion, reaching up to $300$~km/s and extensions of $\sim~2$ and $0.5$~kpc, respectively. They emerge from (or are coincident with) the central regions of these components, being then consistent with the expectations from an outflow scenario. 

\subsection{COS$\_$12306}
This galaxy was identified as one of the most luminous and extended star-forming galaxies at z~$\sim$~6 in the CFHT Legacy Survey \citep{willottExponentialDeclineBright2013}. Being located within the CANDELS field  \citep{groginCANDELSCosmicAssembly2011, koekemoerCANDELSCosmicAssembly2011}, it also has high-resolution HST/ACS and WFC3 imaging available. With an original photo-z of 6.197, it was later spectroscopically confirmed to be at $z\sim5.91$ through the detection of Ly$\alpha$ \citep{pentericciCANDELSz7LargeSpectroscopic2018}, while our observations of the rest-frame optical emission lines provide a lower value, $z\sim5.84$. The system presents a complex morphology in the \OIIIb line emission map with two main components (12306-C, 12306-NW) and several substructures along $\sim$6~kpc, possibly all within a merging system.
The broad component emission seems associated with the brightest central component and extends to the northeast in a structure that is consistent with a conical outflow along the minor axis of the system. 

\subsection{HFLS3}

HFLS3 is a complex, lensed star-forming system at $z=6.3$ with an extreme infrared luminosity of $2.86\times10^{13}$~L$_\odot$ and a SFR of about $2900$~M$_\odot$/yr \citep{riechersDustObscuredMassiveMaximumStarburst2013}. This system has been recently studied in detail with NIRSpec-IFS data by \citet{jonesGANIFSJWSTNIRSpec2024}, who find that instead of a single extreme starburst, the system consists of six star-forming galaxies at $z\sim6.3$ (c1, c2, w1, w2, s1, and s2) that are lensed by two foreground galaxies: g1 ($z\sim3.48$) and g2 ($z\sim2$). Since the spectroscopic setup employed for the R2700 observations (G395H/F290LP) targets the rest-frame optical at $z\sim6.3$, in HFLS3$\_$g1 the only strong optical emission line covered is \Ha, while for HFLS3$\_$g2 the optical emission falls outside the wavelength coverage. 

For the present analysis, we have searched for broad components in all the galaxies within the FoV, finding evidence of the presence of outflows in HFLS3$\_$g1 through \Ha, and in HFLS3$\_$w 1, 2 through \OIII. In particular, for HFLS3$\_$g1 we detect extended emission of $\sim1$~kpc in radius around its centre, with $\sigma_{\rm broad}=100-150$~km/s. We note that broad emission is detected also in forbidden lines, \NIIb and \SIIIline \citep{jonesGANIFSJWSTNIRSpec2024}, confirming the outflow nature of the broad emission. For HFLS3$\_$w 1, 2, we have identified signatures of outflows in the regions around the peaks of continuum emission.

\subsection{CR7}

This complex merging system was first identified by \cite{mattheeIdentificationBrightestLyalpha2015} as one of the most luminous Ly$\alpha$-emitters (L$_{\rm Ly\alpha}\sim 10^{44}$~L$_\odot$) at $z>6$. It consists of three main companion galaxies (CR7$\_$A, CR7$\_$B, CR7$\_$C) and several substructures, all within a projected distance of $\sim8$~kpc. The system has been recently studied using NIRSpec-IFS observations by  \citet{marconciniGANIFSDissectingMultiple2025}, who have probed the star formation history (SFH) of the three main galaxies and the complex interactions among the different components. They also reported the presence of ionised outflows in two galaxies of the system, CR7$\_$A and CR7$\_$B. In our analysis, we obtain similar results. In particular, we identify a broad component extending up to 3~kpc from the emission peak in CR7$\_$A and values of $\sigma_{\rm broad}\sim150-300$~kpc), and a more compact broad region of $<1$~kpc and $\sim200$~km/s around the continuum maximum of CR7$\_$B.

\subsection{COS-3018}

\cite{scholtzGANIFSISMProperties2025} have studied this system with NIRSpec-IFS finding that the SFH, the interstellar medium conditions and kinematics, as well as the stellar population properties are better explained by a merging event of three galaxies (C, NE, SE) than by a rotating disc, as initially inferred from lower resolution ALMA observations \citep{smitRotationIIemittingGas2018}. The main galaxy also shows sub-structure with a set of clumps aligned along the E-W direction, as revealed by NIRCam imaging taken as part of the PRIMER survey (PID1837, PI J. Dunlop).    

\cite{scholtzGANIFSISMProperties2025} already reported the need of fitting the emission lines profiles of the main (C) galaxy with two narrow components and a broad component with $FWHM_{\rm broad}=1210\pm120$~km/s, that they explained as due to a relatively compact outflow ($\sim$~2.1~kpc). To avoid contamination from the NE and SE galaxies, we focus our analysis on a circular aperture with radius 0.15\arcsec (0.8~kpc) at the peak of stellar emission in the C galaxy (see Fig.~\ref{fig:sigma_broad_maps}). In the two-component modelling of the integrated spectrum of this region we identify a broad component with $FWHM_{\rm broad}=787\pm21$~km/s. Although still high, these velocities are lower than those reported by \citet{scholtzGANIFSISMProperties2025}. This difference could be attributed to their use of a larger aperture that includes galaxy C, potentially capturing gas at higher velocities beyond the nuclear regions and to the fact that their spectral modelling includes three kinematic components (two narrow and one broad), which might better account for the high-velocity wings of the emission-line profiles. 

Another broad component seems to be in spatial agreement with the position of the NE galaxy. However, the weak continuum emission in this region does not support a star formation--driven outflow as the cause for the high velocity dispersion, which could instead be produced by the merging of the galaxies.

\subsection{B14}

This massive system consists of two main cores with strong \OIIIb and continuum emission, and other fainter substructures as recently revealed by JWST NIRCam \citep{sugaharaRIOJAComplexDusty2025} and  NIRSpec-IFS \citep{jonesGANIFSWitnessingComplex2025} observations. \cite{jonesGANIFSWitnessingComplex2025} identified the presence of a broad component associated with the integrated spectra of these structures  ($FWHM\sim 550-750$~km/s), as well as an extended broad emission as traced by the $W80$ index over a region of about 4~kpc along the direction of the two main cores. They interpreted these features as due to tidal effects or the presence of an outflow. \citet{prieto-jimenezSpatiallyResolvedHa2025} has recently attempted to identify hints of this broad emission in the \Ha line profile as observed by MIRI, but their analysis did not yield a significant detection. 

As in \cite{jonesGANIFSWitnessingComplex2025}, we detect very extended broad emission beyond the continuum emission and identify regions with red-shifted velocities ($<-200$~kms$^{-1}$) around the W nucleus. Based on these properties, we do not consider 
the presence of a clear outflow in this system, as the broad emission could potentially be a consequence of an ongoing merger or tidal interactions. 

\subsection{MACSJ0416-Y1}

MACSJ0416-Y1 was first selected as a $z=8$ candidate \citep{laporteFrontierFieldsCombining2015} and later on spectroscopically confirmed at $z=8.31$ through the detections of $[\mathrm{O}\textsc{ iii}]$~88~$\mu$m \citep{tamuraDetectionFarinfraredIii2019} and  $[\mathrm{C}\textsc{ ii}]$~157.7~$\mu$m \citep{bakxALMAUncoversII2020}. In a recent study using NIRCam/WFSS data, \citet{maJWSTViewFour2024} estimated high SFRs ($\sim165$M$_\odot$yr$^{-1}$), relatively high dust extinction $A_{\rm V}=0.92-1.10$~mag and very young stellar populations ($<5$~Myr). In our study, we identify a compact ($r_{\rm out}=0.5$~kpc) ionised outflow in a star-forming region at the East of the system, and $\sigma_{\rm out}\sim200$~kms$^{-1}$.

\subsection{EGSY8P7}
This galaxy was initially identified through photometric dropout by \citet[][z$_{phot}$ = 8.57$^{+0.22}_{-0.43}$]{roberts-borsani7GalaxiesRed2016}, and later on confirmed through strong Ly$\alpha$ emission \citep[z$_{spec}$ = 8.683$^{+0.001}_{-0.004}$;][]{zitrinLymanalphaEmissionLuminous2015}, making it one of the most distant Ly$\alpha$ emitters known to date. Recently, NIRSpec MOS medium-resolution observations revealed a tentative detection of a broad \Hb component with $FWHM\sim1200$~kms$^{-1}$ that was interpreted as evidence of a broad-line AGN by \citet{larsonCEERSDiscoveryAccreting2023}. In a detailed analysis using NIRSpec IFS data, \citet{zamoraGANIFSUnderstandingionization2025} identified a broad component in \Hb and \OIIIb with $FWHM\sim650$~kms$^{-1}$, extended over a distance of $\sim1$~kpc and consistent with an outflow driven by stellar feedback. 

In our study, the ionisation in this source does not reveal evidence of AGN activity (see right panels in Fig.~\ref{fig:integrated_line_diagnostics}). We detect an ionised outflow with $r_{\rm out}=1$~kpc and $\sigma_{\rm out}=214.3\pm2.6$~kms$^{-1}$, slightly lower velocities than those reported by \citet{zamoraGANIFSUnderstandingionization2025}.

\subsection{MACS1149-JD1}

MACS1149-JD1 resides behind the MACS J1149.6+2223 cluster and presents a magnification $\mu$~=~$10\pm0.7$ \citep{hashimotoOnsetStarFormation2018a, stiavelliPuzzlingPropertiesMACS1149JD12023, marconciniGANIFSInterplayMerger2024a}. \citet{marconciniGANIFSInterplayMerger2024a}
have recently studied this system with NIRSpec IFS data. They identify a region towards the West with a relatively large velocity dispersion and requiring a second broad component. This is spatially coincident with a peak in the SFR density, and suggests the presence of an outflow. \cite{alvarez-marquezSpatiallyResolvedHa2024} measured the \Ha kinematics with JWST/MIRI suggesting the hypothesis of outflowing gas powered by the UV-bright clump JD1-S. 

Our analysis also supports the presence of a secondary kinematic component towards the West of the system. However, the associated $\sigma_{\rm broad}=68$~km/s is too low to be interpreted as an outflow, and it could be instead a region of overlapping gas between the southern and northern systems. 

\subsection{SPT0311-58}

ALMA observations of the SPT0311-58 system at z~$\sim$~6.9, showed that it resides in a massive dark-matter halo and hosts two dusty galaxies (E and W) with a combined SFR~$\sim$ 3500~M$_{\odot}$/yr \citep{marroneGalaxyGrowthMassive2018b}. Recent NIRSpec-IFS observations have revealed that the system also consists of ten additional smaller galaxies within a region of 17~kpc, representing the core of an extremely massive proto-cluster \citep{arribasGANIFSCoreExtremely2024a}. In this later work, the authors identified a region towards galaxy E that presented an additional kinematic component with a low velocity dispersion ($\sigma<100$~km/s) and redshifted 
velocities ($\sim260$~km/s) with respect to the systemic. Given the relatively high attenuation ($A_{\rm V}=1.7-2.4$) found in galaxy E, a possible outflow in which only the redshifted component was visible was deemed unlikely, and instead this secondary component was interpreted as a satellite galaxy moving towards galaxy E in a minor merger. 

In our analysis, we also find a secondary kinematic component with redshifted velocities although higher velocity dispersion ($\sigma=100-250$~km/s) than those reported by \citet{arribasGANIFSCoreExtremely2024a}. Nevertheless, these results confirm that the additional kinematic component is associated with a satellite.

\section{Kinematic maps}
\label{app:kinematic_maps}
In this section we present, for each system in our sample, the maps of line fluxes and kinematics obtained from the emission-line analysis (of \OIIIb or \Ha) described in Sect.~\ref{subsec:spectral_modelling}. We include maps of integrated line fluxes, velocity and velocity dispersion for the one-component fit (top) and the narrow (middle) and broad (bottom) kinematic components. These maps, in particular the ones displaying the velocity and velocity dispersion of the broad component, are used to define the regions hosting ionised outflows, which are highlighted in red, as in Fig.~\ref{fig:sigma_broad_maps}. The maps are presented as supplementary material at \url{https://doi.org/10.5281/zenodo.18032864}.

\section{Spectral modelling of outflow spectra}
\label{app:outflow_fits}
This section contains the spectral modelling of the integrated spectrum of the outflows identified in our sample of star-forming systems, including the main emission lines identified in each case. When an emission line is better modelled with two kinematic components (see Sect.~\ref{sec:outflow_regions}) we show narrow (blue) and broad (red) components, otherwise we  only show the one-component model (green). The figures with the spectral modelling of each target are presented as supplementary material at \url{https://doi.org/10.5281/zenodo.18032864}.

 \end{appendix}

\end{document}